\documentclass[aps,pra,reprint,10pt,twocolumn,superscriptaddress,floatfix]{revtex4-1}
\usepackage{times}
\usepackage{amsmath,amssymb,bm,graphicx}
\usepackage[breaklinks,colorlinks=true,urlcolor=blue,citecolor=blue,linkcolor=blue]{hyperref}
\usepackage{mathtools}
\usepackage{xcolor}
\usepackage{enumitem}
\usepackage{braket}
\usepackage{xspace}
\usepackage{grffile}
\usepackage[caption=false]{subfig}
\usepackage{bbm}

\definecolor{ourgreen}{rgb}{0,0.5,0}
\definecolor{hzred}{rgb}{0.75,0.3,0.5}

\begin{document}

\title{Hierarchical Clifford transformations to reduce entanglement in quantum chemistry wavefunctions}

\date{\today}

\author{Ryan V. Mishmash}
\email{ryan.mishmash@gmail.com}
\altaffiliation[Current affiliation: ]{Microsoft Station Q}
\affiliation{IBM Quantum, Almaden Research Center, San Jose, California 95120, USA}

\author{Tanvi P. Gujarati}
\affiliation{IBM Quantum, Almaden Research Center, San Jose, California 95120, USA}

\author{Mario Motta}
\affiliation{IBM Quantum, Almaden Research Center, San Jose, California 95120, USA}

\author{Huanchen Zhai}
\affiliation{Division of Chemistry and Chemical Engineering, California
Institute of Technology, Pasadena, California 91125, United States}

\author{Garnet Kin-Lic Chan}
\affiliation{Division of Chemistry and Chemical Engineering, California
Institute of Technology, Pasadena, California 91125, United States}

\author{Antonio Mezzacapo}
\email{mezzacapo@ibm.com}
\affiliation{IBM Quantum, T. J. Watson Research Center, Yorktown Heights, NY 10598, USA}

\begin{abstract}
The performance of computational methods for many-body physics and chemistry is strongly dependent on the choice of basis used to cast the problem; hence, the search for better bases and similarity transformations is important for progress in the field. So far, tools from theoretical quantum information have been not thoroughly explored for this task. Here we take a step in this direction by presenting efficiently computable Clifford similarity transformations for quantum chemistry Hamiltonians, which expose bases with reduced entanglement in the corresponding molecular ground states. These transformations are constructed via block diagonalization of a hierarchy of truncated molecular Hamiltonians, preserving the full spectrum of the original problem. We show that the bases introduced here allow for more efficient classical and quantum computation of ground state properties. First, we find a systematic reduction of bipartite entanglement in molecular ground states as compared to standard problem representations. This entanglement reduction has implications in classical numerical methods such as ones based on the density matrix renormalization group. Then, we develop variational quantum algorithms that exploit the structure in the new bases, showing again improved results when the hierarchical Clifford transformations are used.
\end{abstract}

\maketitle

\section{Introduction}

Similarity transformations (STs) have long been used to support the investigation of  many-particle quantum systems. Ideally, a good similarity transformation is one that can be obtained efficiently (computable in polylogarithmic time in the Hilbert space size), preserves the full spectrum of the original Hamiltonian, and is unitary---this last requirement being specifically important for quantum computations.

Because of their ubiquity, similarity transformation techniques are known by a variety of names. For example, Schrieffer-Wolff (SW) transformations are a classic type of ST \cite{roussy1973approximate,cederbaum1989block,bravyi2011schrieffer,white2002numerical} used to obtain approximate representations of low-energy levels by decoupling high-energy degrees of freedom. In the context of electronic structure, STs are the starting point for several methods to solve the Schr\"{o}dinger equation, including the coupled cluster~\cite{bartlett2007coupled} and canonical transformation approaches~\cite{white2002numerical,yanai2006canonical,neuscamman2010review}, and have been used to accelerate basis set convergence in quantum chemistry.
In the above cases, the most general form of the ST usually cannot be implemented efficiently. Consequently, restrictions are applied, for example, by approximating the similarity transformed Hamiltonian with an expansion truncated at finite order, or by restricting the form of the ST to low-particle rank excitations. 

Recent years have witnessed research at the interface between STs, quantum chemistry, and quantum computation. These include, for example, the construction of active-space Hamiltonians for quantum computing simulations~\cite{bauman2019downfolding,bauman2019quantum,metcalf2020resource} and the combination of quantum computing algorithms with the transcorrelated method to transfer explicit correlation from the wavefunction to a similarity-transformed non-Hermitian Hamiltonian~\cite{mcardle2020improving,motta2020quantum,sokolov2022orders}, or a
canonically transformed Hermitian Hamiltonian~\cite{kumar2022quantum}.
%More generally, 
The STs used above 
%from the incorporation of quantum computing subroutines in the workflow of 
correspond to the existing classes of STs used in electronic structure. Therefore, it is a natural question whether new classes of STs can be designed---based instead on concepts from quantum information science---and suitably adapted to the domain of quantum chemistry.

In this work, we define Clifford STs based on the stabilizer formalism~\cite{gottesman_stabilizer_1997, bravyi_tapering_2017}.
As they lie in the Clifford group, they can be used to exactly and efficiently transform the many-body Hamiltonian. 
%which reveal the structure of quantum chemistry models. 
We describe an efficient classical procedure to define a set of Clifford STs
%construct a set of unitary STs, at efficient classical computational cost, 
from a series of truncated representations of the second-quantized electronic many-body Hamiltonian as a multi-qubit operator. These STs reveal a hierarchy of hidden approximate binary symmetries and entanglement patterns in the structure of the Hamiltonian, and through these approximate symmetries the STs lead to approximate decouplings of the quantum degrees of freedom. 
While the numerical procedure to determine the Clifford STs involves `truncations' of the multi-qubit Hamiltonian, it is worth stressing that the STs do not change the Hamiltonian spectrum. The resulting (approximately) decoupled similarity transformed Hamiltonian can thus be used to improve the cost and accuracy of classical and quantum simulations.

The results in this manuscript are organized as follows. We revisit a procedure to remove (`taper') qubits from Hamiltonians based on $\mathbb{Z}_2$ Pauli symmetries in Sec.~\ref{sec:z2review}, adapting the previously established formalism for the purpose of block diagonalization. In Sec.~\ref{sec:hct}, we present the main result, showing that a hierarchy of block-diagonalizing Clifford STs can arise from new symmetries that appear in chemistry Hamiltonians if one truncates them at different thresholds. We perform numerical calculations of the number of new symmetries appearing in truncated Hamiltonians in Sec.~\ref{sec:sym_analysis}. We then investigate the implications of these newfound STs for the entanglement structure of quantum chemical ground states in Sec.~\ref{sec:entanglement}, and for the variational quantum eigensolver in Sec.~\ref{sec:vqe}. Finally, we discuss our results and give an outlook of the potential for future use of our hierarchical Clifford STs for both classical and quantum computations in Sec.~\ref{sec:conclusion}.

\section{Pauli symmetries and \\ block-diagonal Hamiltonians} \label{sec:z2review}

We start by casting the full second-quantized electronic many-body Hamiltonian,
\begin{equation}
H_\mathrm{elec} = E_0 + \sum_{ \substack{pq\\\sigma} } h_{pq} \, \hat{c}_{p\sigma}^\dagger 
\hat{c}_{r\sigma}^{\phantom{\dagger}}
+ 
\sum_{ \substack{prqs\\\sigma\tau} }
\frac{(pr|qs)}{2} \, 
\hat{c}_{p\sigma}^\dagger 
\hat{c}_{q\tau}^\dagger 
\hat{c}_{s\tau}^{\phantom{\dagger}}
\hat{c}_{r\sigma}^{\phantom{\dagger}} \, ,
\label{eq:Helec}
\end{equation}
into the form of a real linear combination of $n$-qubit Pauli operators:
\begin{equation}
    H = \sum_{i=1}^{N_\mathrm{Pauli}} h_i P_i,
    \label{eq:Hpauli}
\end{equation}
with $h_i \in \mathbb{R}$ and $P_i \in \{I, \sigma^x, \sigma^y, \sigma^z\}^{\otimes n}$, via a prescribed qubit mapping~\cite{bravyi_fermionic_2002, bravyi_tapering_2017} (e.g., Jordan-Wigner, Bravyi-Kitaev, or parity mapping). 
We can efficiently identify a symmetry group $\mathcal{S} = \langle \tau_1, \dots, \tau_{n_\mathrm{sym}} \rangle \subseteq \mathcal{P}_n = \pm\{I, \sigma^x, \sigma^y, \sigma^z\}^{\otimes n}$ with $-I\not\in\mathcal{S}$ such that all elements of $\mathcal{S}$ commute with \emph{each} Pauli term in $H$: $[\sigma, P_i] = 0~\forall~\sigma\in\mathcal{S},\,i=1,\dots,N_\mathrm{Pauli}$.

In order to find the generators $\tau_1, \dots, \tau_{n_\mathrm{sym}}$,
we represent Pauli operators as binary strings $(a_x|a_z)$. A Pauli operator is then recovered via $\boldsymbol{\sigma}(a_x|a_z) =e^{i\phi} ( \prod_{i\in a_x} \sigma^x_i ) \cdot ( \prod_{j\in a_z} \sigma^z_j )$. 
for some phase factor $e^{i\phi}$ that we shall ignore. 
We represent the qubit Hamiltonian $H$ as a binary matrix 
\[
G=\left[ \begin{array}{c}
G_x | G_z 
\end{array} \right]
\]
where $\boldsymbol{\sigma}_j=\boldsymbol{\sigma}(G_x^j|G_z^j)$.
The symmetry generators of the group $\cal{S}$ are then efficiently found by solving for the generating basis of the binary kernel of $G$. This is because in this binary strings representation commutation between two Pauli operators amounts to a binary product. 

Eigenstates of $H$ can thus be labeled by eigenvalues of the set of $n_\mathrm{sym}\equiv |\mathcal{S}|$ independent generators $\{\tau_j\}$. Each generator $\tau_j$ corresponds to a $\mathbb{Z}_2$ Pauli symmetry with parity eigenvalues $\pm1$, and thus the Hamiltonian can be organized into $2^{n_\mathrm{sym}}$ blocks each labeled by a set of eigenvalues from $\{+1,-1\}^{n_\mathrm{sym}}$. Choosing a set of $n_\mathrm{sym}$ fixed parity eigenvalues thereby allows one to `taper' the corresponding symmetry qubits~\cite{bravyi_tapering_2017}, which is now a well-established technique in variational quantum chemistry.

This block-diagonal structure is made particularly transparent in the qubit language where (owing to $\mathcal{S}$ being a stabilizer group~\cite{gottesman_stabilizer_1997}) we can explicitly construct a unitary Clifford operator
\begin{equation}
    C_\mathrm{tapering} = \prod_{j=1}^{n_\mathrm{sym}} C_j,
    \label{eq:Cfull}
\end{equation}
with
\begin{equation}
    C_j = \frac{1}{\sqrt{2}}(\sigma_{q(j)} + \tau_j),
    \label{eq:Cj}
\end{equation}
which transforms each generator $\tau_j$ to a corresponding \emph{single-qubit} Pauli operator $\sigma_{q(j)}$ acting only on qubit $q(j)$:
\begin{equation}
    C_\mathrm{tapering}^\dagger \tau_j C_\mathrm{tapering} = \sigma_{q(j)},\quad j = 1, \dots, n_\mathrm{sym}.
\end{equation}
The operators $C_j$ are constructed such that the single-qubit operator $\sigma_{q(j)}$ anticommutes with $\tau_j$ but commutes with $\tau_{j'\neq j}$~\cite{bravyi_tapering_2017}:
\begin{equation}
    \sigma_{q(j)} \tau_{j'} = (-1)^{\delta{_{j,j'}}} \tau_{j'} \sigma_{q(j)}.
    \label{eq:sigmataucomm}
\end{equation}

For simplicity, we assume from here on that the symmetry generators $\tau_j$ are $Z$-type Pauli operators (which is the case in all examples considered herein and in Ref.~\cite{bravyi_tapering_2017}) such that the single-qubit operators $\sigma_{q(j)}$ representing the symmetries in the transformed basis are of $X$-type: $\sigma_{q(j)} \to \sigma^x_{q(j)}$. After conjugating the Hamiltonian by the full Clifford $C_\mathrm{tapering}$,
\begin{equation}
    H' = C_\mathrm{tapering}^\dagger H C_\mathrm{tapering} = \sum_i h_i P'_i,
    \label{eq:Hp}
\end{equation}
where $P'_i \equiv C_\mathrm{tapering}^\dagger P_i C_\mathrm{tapering} \in \mathcal{P}_n$, the single-qubit operators $\sigma^x_{q(j)}$ (transformed generators) now commute with each term in $H'$: $[\sigma^x_{q(j)}, P'_i] = 0$. That is, the subset of qubits $\{q(j)\}$ of every Pauli string $P'_i$ must be an $X$-type Pauli operator, i.e., a product of only identity and $\sigma^x$ single-qubit operators. The ground state of $H'$ then takes the general form
\begin{equation}
    \ket{\psi} = \left[\prod_{j=1}^{n_\mathrm{sym}} \ket{\pm}_{q(j)}\right] \otimes \ket{\Psi},
    \label{eq:Psifull}
\end{equation}
with $\sigma^x_{q(j)}\ket{\pm}_{q(j)} = \pm \ket{\pm}_{q(j)}$. The first factor can be thought of as labeling the block in the block-diagonal Hamiltonian $H'$ [Eq.~\eqref{eq:Hp}], while the second factor is the ground state of that block which acts only on $n-n_\mathrm{sym}$ qubits. (Note that the $n_\mathrm{sym}$ `symmetry qubits' do not factorize out at the level of operators in $H'$; rather the eigenstates must result in a factorized form due to the block-diagonal structure of the Hamiltonian.)

As mentioned above, it is customary to now `taper' off $n_\mathrm{sym}$ qubits in $H'$ by replacing $\sigma^x_{q(j)}$ in the $P'_i$ with the respective parity eigenvalues $\{+1,-1\}^{n_\mathrm{sym}}$ in a fixed symmetry sector and solve for the ground state of the remaining ($n-n_\mathrm{sym}$)-qubit problem~\cite{bravyi_tapering_2017}. For chemistry problems, the sector is typically chosen to coincide with that of the Hartree-Fock ground state, but in general the global ground state may lie in any of the allowed $2^{n_\mathrm{sym}}$ sectors. In fact, below in Sec.~\ref{sec:vqe} we develop VQE schemes for the `truncated' models discussed in the next section for which the sector of the ground state is not known or cannot be guessed a priori and thus needs to be determined dynamically in the optimization.

\section{Hierarchical Clifford Transformations} \label{sec:hct}

In this section, we generalize the Clifford ST described in the previous section to construct new Clifford STs based on a \emph{hierarchical decoupling} scheme of the multi-qubit Hamiltonian, Eq.~\eqref{eq:Hpauli}. Below, we describe such \emph{hierarchical Clifford transformations} (HCTs) proper, but as a warm up we first spell out important primitives of the construction in detail. Given an `energy threshold' $\epsilon > 0$, we define the corresponding truncated Hamiltonian as follows:
\begin{equation}
    H_{\epsilon} = \sum_{i:\,|h_i| \geq \epsilon} h_i P_i.
    \label{eq:Heps0}
\end{equation}
Since the generators $\{\tau_j\}$ of the symmetry group $\mathcal{S}$ identified in Sec.~\ref{sec:z2review} commute with all Pauli terms in the full model $H$ separately and $H_{\epsilon}$ merely represents an elimination of certain Pauli terms from $H$, it is possible to identify a symmetry group $\mathcal{S}_\epsilon = \langle \tau^{\epsilon}_1, \dots, \tau^{\epsilon}_{n_\epsilon} \rangle$ of $H_{\epsilon}$ such that $\mathcal{S} \subseteq \mathcal{S}_\epsilon$ (and thus $n_\epsilon \geq n_\mathrm{sym}$).
However, naively applying the algorithm from Ref.~\cite{bravyi_tapering_2017} to $H_{\epsilon}$ (as implemented in Qiskit's \texttt{Z2Symmetries} class~\cite{qiskit}) will not generally satisfy this subset relation \emph{at the level of the generators}, which is important for practical use as we will see later.
% We are also careful to choose a representation of the group $\mathcal{S}_0$ such that the sets of generators themselves obey the same relation:
% $\{\tau_j\}_{j=1,\dots,n_\mathrm{sym}} \subseteq \{\tau^{(0)}_j\}_{j=1,\dots,n_0}$.

In order to ensure that we choose a representation of $\mathcal{S}_\epsilon$ with $\{\tau_j\}_{j=1,\dots,n_\mathrm{sym}} \subseteq \{\tau^{\epsilon}_j\}_{j=1,\dots,n_\epsilon}$, we employ the following procedure. We first identify generators $\{\tau_j\}$ (`symmetries') and corresponding single-qubit operators $\{\sigma^x_{q(j)}\}$ (`symmetry qubits') for $H$ following Ref.~\cite{bravyi_tapering_2017}. We then apply the same (naive) algorithm to $H_{\epsilon}$ to obtain $n_\epsilon$ `bare' symmetries $\{\tilde{\tau}^{\epsilon}_j\}$ and symmetry qubits $\{\sigma^x_{q_\epsilon(j)}\}$ \footnote{Since we always have $X$-type Pauli operators for the symmetry qubits also for the truncated model, we can take $\sigma^{\epsilon}_{q_\epsilon(j)} \to \sigma^x_{q_\epsilon(j)}$.}, the latter of which live on a set of qubit indices $\{q_\epsilon(j)\}_{j=1,\dots,n_\epsilon} \supseteq \{q(j)\}_{j=1,\dots,n_\mathrm{sym}}$. Next, we identify the \emph{added} symmetry qubits as the following set difference: $\Delta q_{0,\epsilon} \equiv \{q_\epsilon(j)\}_{j=1,\dots,n_\epsilon}\setminus\{q(j)\}_{j=1,\dots,n_\mathrm{sym}}$. Finally, we can obtain a final set of $n_\epsilon$ independent generators for $\mathcal{S}_\epsilon$ which explicitly contains the original symmetries of $H$:
\begin{equation}
    \left\{\tau^{\epsilon}_j\right\} = \left\{\tau_j\right\} \cup \left\{\tau^{0,\epsilon}_j\right\}
    \label{eq:tauepsj},
\end{equation}
where
\begin{equation}
    \left\{\tau^{0,\epsilon}_j\right\} = \left\{\tilde{\tau}^{\epsilon}_j\right\}_{j \in \{j' | q_\epsilon(j') \in \Delta q_{0,\epsilon}\}},
    \label{eq:tau0epsj}
\end{equation}
and the set of associated symmetry qubit operators can be taken to be the $\{\sigma^x_{q_\epsilon(j)}\}$. Ultimately, this procedure provides an efficient means of finding $n_\epsilon = |\mathcal{S}_\epsilon|$ symmetries of $H_{\epsilon}$: Of these, $n_\mathrm{sym}$ ($n_\epsilon - n_\mathrm{sym}$) are exact (approximate) symmetries of the full model $H$; i.e., the first and second sets, respectively, in the set union on the right-hand-side of Eq.~\eqref{eq:tauepsj} (and similarly for the symmetry qubit operators).

The Clifford unitary that block diagonalizes the truncated Hamiltonian (and thus `approximately block diagonalizes' the full Hamiltonian) is given by
\begin{equation}
    C(\epsilon) = C_\mathrm{tapering} C_{0,\epsilon},
    \label{eq:Ceps}
\end{equation}
where
\begin{equation}
    C_{0,\epsilon} = \prod_{j=1}^{n_\epsilon - n_\mathrm{sym}} C^{0,\epsilon}_j
    \label{eq:C0eps}
\end{equation}
and
\begin{equation}
    C^{0,\epsilon}_j = \frac{1}{\sqrt{2}}\left(\sigma^x_{\Delta q_{0,\epsilon}(j)} + \tau^{0,\epsilon}_j \right).
    \label{eq:C0epsj}
\end{equation}
We stress that the algorithm for finding this transformation is \emph{efficient} (polynomial) as it uses the same stabilizer-based machinery as that used to find exact symmetries of $H$ summarized in the previous section.

While the single-qubit operator and symmetry generator present on the right-hand-side of Eq.~\eqref{eq:C0epsj} anticommute (for the same $j$), i.e.,
\begin{equation}
    \left[\sigma^x_{\Delta q_{0,\epsilon}(j)}, \tau^{0,\epsilon}_j\right]_{+} = 0,
    \label{eq:sigmatauanticommeps}
\end{equation}
the commutation relations analogous to those in Eq.~\eqref{eq:sigmataucomm} now read
\begin{equation}
    \left[\sigma^{x}_{q(j')}, \tau^{0,\epsilon}_{j}\right]_{-} = 0,
    \label{eq:sigmataucommeps}
\end{equation}
for all $j = 1,\dots,n_{\epsilon}-n_\mathrm{sym}$ and $j' = 1,\dots,n_\mathrm{sym}$. In other words, the identified approximate symmetries, $\left\{\tau^{0,\epsilon}_{j}\right\}$,  must commute with the already-identified (exact) symmetry qubits, $\left\{\sigma^{x}_{q(j')}\right\}$. If this condition were not satisfied, the additional Clifford $C_{0,\epsilon}$ in Eq.~\eqref{eq:Ceps} would spoil the block-diagonal structure resulting from the exact symmetries as implemented in the ST by $C_\mathrm{tapering}$.

\begin{figure}[t]
  \begin{center}
  \includegraphics[width=0.8\columnwidth]{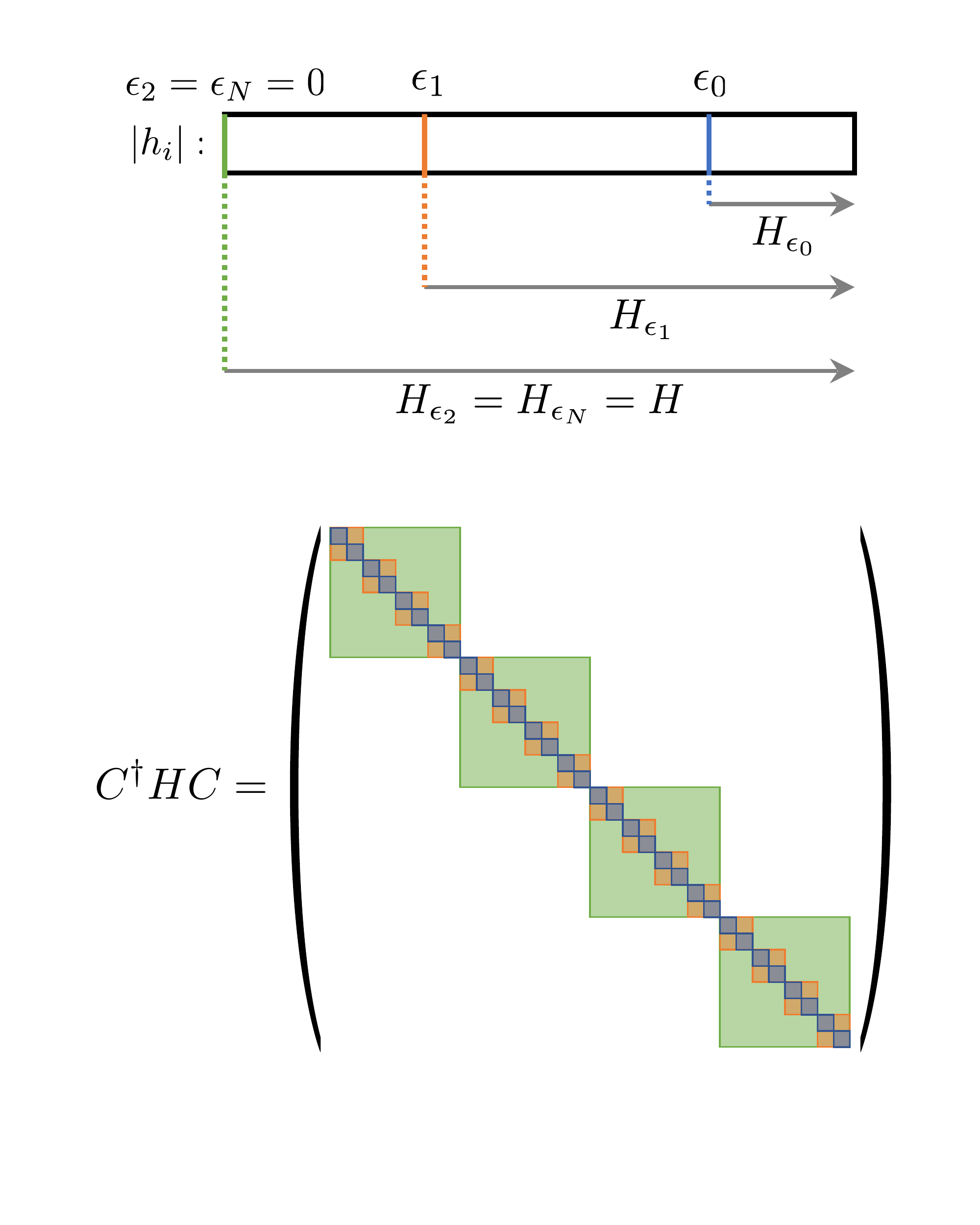}
  \end{center}
  \caption{(top) Schematic of the $H_{\epsilon_m}$ defined by truncating the full Hamiltonian $H = \sum_i h_i P_i$ according to the absolute values of the coefficients $|h_i|$ [cf.~Eq.~\eqref{eq:Hepsm}] for a minimal example of $N=2$ thresholds: $\epsilon_2 = \epsilon_N = 0 < \epsilon_1 < \epsilon_2$; $|h_i|$ is represented as increasing from left to right in the diagram. (bottom) Schematic of a ($2^n \times 2^n$) matrix representation of the untruncated Hamiltonian, $H$, conjugated by the corresponding hierarchical Clifford unitary, $C = C(\{\epsilon_m\})$, as described in Sec.~\ref{sec:hct}. In this hypothetical example, we have $n_\mathrm{sym}=2$ exact symmetries ($2^2=4$ blocks; green) at threshold level $\epsilon_2 = \epsilon_N = 0$, $n_1=4$ (2 approximate, 2 exact) symmetries ($2^4=16$ blocks; orange) derived from $\epsilon_1$, and $n_2=5$ (3 approximate, 2 exact) symmetries ($2^5=32$ blocks; blue) derived from the highest threshold $\epsilon_0$. Matrix elements with larger magnitude live within smaller blocks by construction.
  \label{fig:Fig1}
  }
\end{figure}

Under the Clifford ST generated by $C(\epsilon)$ the full Hamiltonian, $H = H_{\epsilon} + \Delta H_{\epsilon}$, reads
\begin{equation}
    C^\dagger(\epsilon) H C(\epsilon) = C^\dagger(\epsilon) H_{\epsilon} C(\epsilon) + C^\dagger(\epsilon) \Delta H_{\epsilon} C(\epsilon).
    \label{eq:Htransformed}
\end{equation}
The first term is block diagonal on the $n_\epsilon$ qubits encoding the $n_\epsilon$ symmetries of $H_{\epsilon}$, while the second term encodes off-block-diagonal elements of $H$ (cf.~Fig.~\ref{fig:Fig1}).
%Because $n_0$ can be significantly larger than $n_\mathrm{sym}$, the former problem can be substantially simpler.

To what extent are the exact symmetries of $H_{\epsilon}$ only approximate symmetries of the full $H$? A straightforward application of the triangle inequality gives the following bound in terms of transformed operators (recall that $[\sigma^x_{q_\epsilon(j)}, C^\dagger(\epsilon) H_{\epsilon} C(\epsilon)] = 0$ by construction):
\begin{equation}
    \left\|[\sigma^x_{q_\epsilon(j)}, C^\dagger(\epsilon) H C(\epsilon)]\right\| \leq \sum_{i:\,|h_i| < \epsilon} 2|h_i|,
    \label{eq:symerrorbound}
\end{equation}
where $||\cdot||$ denotes the operator norm. It is also straightforward to show---given the important commutation relation in Eq.~\eqref{eq:sigmataucommeps}---that the single-qubit Pauli operators representing the symmetries of the full model, i.e., the $\{\sigma^x_{q(j)}\}_{j=1,\dots,n_\mathrm{sym}} \subseteq \{\sigma^x_{q_\epsilon(j)}\}_{j=1,\dots,n_\epsilon}$ defined in Sec.~\ref{sec:z2review}, (naturally) remain \emph{exact} symmetries of $H$ after conjugation by $C(\epsilon)$: $[\sigma^x_{q(j)}, C^\dagger(\epsilon) H C(\epsilon)] = 0$.

While the Clifford ST defined in Eq.~\eqref{eq:Ceps} formally represents an efficient means of automatically and efficiently transforming a Hamiltonian $H$ to an approximately block-diagonal form, it may be of limited practical use. Ideally, we would like to maximize the total number of symmetries $n_\epsilon$ in $H_{\epsilon}$.
%in order to, say, most easily obtain a solution to the (block-diagonal) starting point $C_0 H_{\epsilon_0} C_0^\dagger$.
However, this will generally require a relatively large $\epsilon$ such that the additional symmetries gained by decoupling (i.e., the ultimately approximate symmetries of $H$)
%$\{\tilde{\tau}^{(0)}_j\}_{j \in \{j' | q_0(j') \in \Delta q_0\}}$ from Eq.~\eqref{eq:tau0j}]
may not be useful for solution of $H$ itself. As an example, consider an $H_{\epsilon}$ obtained by truncating $H$ such that only $Z$-type Pauli operators remain, i.e., the truncated model is completely classical with a corresponding ground state in the form of a single computational basis state (the Hartree-Fock bitstring is one such state). In this case, the algorithm described above will produce a set of generators $\{\tau^{\epsilon}_j\}$ for $\mathcal{S}_\epsilon$ consisting of (1) the $n_\mathrm{sym}$ exact symmetries of $H$, i.e, $\{\tau_j\}$, and (2) $n_\epsilon - n_\mathrm{sym}$ approximate symmetries [Eq.~\eqref{eq:tau0epsj}] which can be chosen to consist entirely of \emph{single-qubit} Pauli-$Z$ operators.
% [i.e., the corresponding Clifford operators in Eq.~\eqref{eq:C0epsj} are purely \emph{local}].
The latter are a relatively trivial manifestation of aggressive truncation and so it would be desirable to identify a Clifford ST for which the approximate symmetries have more meaningful and useful physical structure---indeed if the identified approximate symmetries are purely local, $C_{0,\epsilon}$ will be a \emph{local} Clifford ST which by definition cannot alter the entanglement structure of the model in the many-qubit basis (see Sec.~\ref{sec:entanglement} below).

The Clifford unitary $C(\epsilon)$ described above can in fact be generalized by using a hierarchical scheme, thereby generating the advertised \emph{hierarchical Clifford transformations} (HCTs). Instead of defining merely a single threshold $\epsilon$, we consider a hierarchical sequence of $N$ (finite) energy thresholds $\epsilon_{m=0,...,N-1} > 0$ with $\epsilon_{m+1} < \epsilon_m$ and concomitant energy-truncated Hamiltonians as in Eq.~\eqref{eq:Heps0}:
\begin{equation}
    H_{\epsilon_m} = \sum_{i:\,|h_i| \geq \epsilon_m} h_i P_i.
    \label{eq:Hepsm}
\end{equation}
We also define $\epsilon_N \equiv 0$ such that $H_{\epsilon_N} \equiv H$. For a schematic of this series of Hamiltonians, see the top panel of Fig.~\ref{fig:Fig1}.

The Clifford transformation analogous to Eq.~\eqref{eq:Ceps} based on the full schedule of $N+1$ thresholds $\{\epsilon_m\}_{m=0,\dots,N}$ (including $\epsilon_N \equiv 0$) reads as follows:
\begin{equation}
    C(\{\epsilon_m\}) = \prod_{i=0}^N C_{\epsilon_{N-i+1}\epsilon_{N-i}},
    \label{eq:Chct}
\end{equation}
where, analogous to Eqs.~\eqref{eq:C0eps}-\eqref{eq:C0epsj},
\begin{equation}
    C_{\epsilon_{m+1}\epsilon_{m}} = \prod_{j=1}^{n_{\epsilon_{m}}-n_{\epsilon_{m+1}}} C^{\epsilon_{m+1}\epsilon_{m}}_j,
    \label{eq:Cepsmp1epsm}
\end{equation}
and
\begin{equation}
    C^{\epsilon_{m+1}\epsilon_{m}}_j = \frac{1}{\sqrt{2}}\left(\sigma^x_{\Delta q_{\epsilon_{m+1}\epsilon_{m}}(j)} + \tau^{\epsilon_{m+1}\epsilon_{m}}_j \right).
    \label{eq:Cepsmp1epsmj}
\end{equation}
The first term in Eq.~\eqref{eq:Cepsmp1epsmj} represents the `added' symmetry qubits' in $H_{\epsilon_m}$ relative to $H_{\epsilon_{m+1}}$, while the second term represents the associated `added' symmetry generators, as per the discussion surrounding Eq.~\eqref{eq:tau0epsj}. Specifically, we have
\begin{equation}
    \Delta q_{\epsilon_{m+1},\epsilon_{m}} = \{q_{\epsilon_m}(j)\}_{j=1,\dots,n_{\epsilon_m}}\setminus\{q_{\epsilon_{m+1}}(j)\}_{j=1,\dots,n_{\epsilon_{m+1}}}
    \label{eq:Deltaqepsmp1epsm}
\end{equation}
and
\begin{equation}
    \left\{\tau^{\epsilon_{m+1}\epsilon_{m}}_j\right\} = \left\{\tilde{\tau}^{\epsilon_m}_j\right\}_{j \in \{j' | q_{\epsilon_m}(j') \in \Delta q_{\epsilon_{m+1},\epsilon_{m}}\}},
    \label{eq:tauepsmp1epsmj}
\end{equation}
where again $\{q_\epsilon(j)\}$ and $\{\tilde{\tau}^\epsilon_j\}$ represent the `bare' symmetry qubits and generators of $H_\epsilon$ obtained by applying the algorithm of Ref.~\cite{bravyi_tapering_2017}, as summarized in Sec.~\ref{sec:z2review}, directly to $H_\epsilon$. % disregarding symmetries already found at lower thresholds.

Importantly, we construct the full hierarchical Clifford ST---$C(\{\epsilon_m\})$---by `starting at' zero threshold, i.e., the first factor in the product of Eq.~\eqref{eq:Chct}, is $C_{\epsilon_{N+1}\epsilon_{N}} \equiv C_\mathrm{tapering}$, and the remainder of the transformation is built by identifying additional approximate symmetries via successively \emph{increasing} the threshold. Also, if $n_{\epsilon_m} = n_{\epsilon_{m+1}}$ in Eq.~\eqref{eq:Cepsmp1epsm}, then there have been no additional symmetries identified at $\epsilon_{m}$ relative to $\epsilon_{m+1}$, and $C_{\epsilon_{m+1}\epsilon_{m}} = \mathbbm{1}$ [in this case, Eqs.~\eqref{eq:Deltaqepsmp1epsm} and \eqref{eq:tauepsmp1epsmj} are empty sets]. This is the origin of the `plateaus' appearing in the plots of number of symmetries versus threshold in Figs.~\ref{fig:small_molecules_pople}, \ref{fig:small_molecules_basis}, and \ref{fig:vqe_syms_vs_epsilon} below. Finally, similar to Eqs.~\eqref{eq:sigmatauanticommeps} and \eqref{eq:sigmataucommeps}, we have the following (anti-)commutation relations:
\begin{equation}
    \left[\sigma^x_{\Delta q_{\epsilon_{m+1}\epsilon_{m}}(j)}, \tau^{\epsilon_{m+1}\epsilon_{m}}_j\right]_{+} = 0
\end{equation}
for $j = 1,\dots,n_{\epsilon_{m}}-n_{\epsilon_{m+1}}$ at each $m=N,\dots,0$, and 
\begin{equation}
    \left[\sigma^{x}_{q_{\epsilon_{m+1}}(j')}, \tau^{\epsilon_{m+1}\epsilon_{m}}_{j}\right]_{-} = 0,
\end{equation}
for all $j = 1,\dots,n_{\epsilon_{m}}-n_{\epsilon_{m+1}}$ and $j' = 1,\dots,n_{\epsilon_{m+1}}$.

The full Hamiltonian conjugated by this Clifford unitary, i.e., $C^\dagger(\{\epsilon_m\}) H C(\{\epsilon_m\})$, can be represented [with a particular choice of the tensor product space, cf.~Eq.~\eqref{eq:Psifull}] as an \emph{approximately block-diagonal} matrix as illustrated in the bottom panel of Fig.~\ref{fig:Fig1}. One can equivalently view the full Clifford ST generated by $C(\{\epsilon_m\})$ as consisting of a composition of a hierarchy of Clifford STs which respectively exactly block diagonalize the series of truncated Hamiltonians $H_{\epsilon_m}$; thus, hereafter we refer to the full $C(\{\epsilon_m\})$ as generating a `hierarchical Clifford transformation' (HCT). For the $n_{\epsilon_0} - n_\mathrm{sym}$ identified approximate symmetries, the extent to which they deviate from being exact symmetries of $H$ can be formally bounded in a similar fashion as in Eq.~\eqref{eq:symerrorbound}. Specifically, for the transformed operators, we have
\begin{equation}
    \left\|[\sigma^x_{q_{\epsilon_{m'}}(j)}, C^\dagger(\{\epsilon_m\}) H C(\{\epsilon_m\})]\right\| \leq \sum_{i:\,|h_i| < \epsilon_{m'}} 2|h_i|.
    \label{eq:symerrorbound_general}
\end{equation}
for symmetries identified at threshold $\epsilon_{m'}$.

It is important to note that the set of generators obtained at the final step, i.e., $\{\tau^{\epsilon_0}_j\}$, now depend on the details of the truncation schedule, i.e., the chosen $\epsilon_{m=0,\dots,N-1}$, and not merely on $\epsilon_0$ itself [as was the case in Eq.~\eqref{eq:tauepsj}]. In particular, for an infinitely finely spaced grid ($\epsilon_{m} - \epsilon_{m+1} \to 0^+$), the final set of obtained generators $\{\tau^{\epsilon_0}_j\}_{j=1,\dots,n_{\epsilon_0}}$ consists of \emph{all} encountered symmetries of the set of truncated Hamiltonians $H_{\epsilon}$ for $0 < \epsilon \leq \epsilon_0$ (as well as, of course, the $n_\mathrm{sym}$ exact symmetries of $H = H_{\epsilon_N=0}$ from Sec.~\ref{sec:z2review}: $\{\tau^{\epsilon_N}_j\} = \{\tau_j\}$). In this case, the full HCT is defined entirely by the largest threshold $\epsilon_0$, and it is this ST (with $\epsilon_{m} - \epsilon_{m+1} \to 0^+$) that is used in the analysis in Sec.~\ref{sec:entanglement}. Naturally, $\epsilon_0$ should be chosen large enough such that $n_{\epsilon_0}=n$ (the number of qubits in the system) in order to give the most structure (cf.~bottom panel of Fig.~\ref{fig:Fig1}); indeed, the largest number of (non-identity) factors in Eq.~\eqref{eq:Chct} is $n$.

Equations~\eqref{eq:Chct}-\eqref{eq:tauepsmp1epsmj} constitute the main formal result of this work. The identified HCT can be viewed as \emph{automatically, efficiently, and optimally} identifying approximate symmetries of $H$ and bringing it into a suitable \emph{approximately block-diagonal} form in the local (Pauli $\sigma^x$) basis of qubits. In the following sections, we investigate the implications of this family of STs for classical and quantum simulation of quantum chemical systems.

\section{Dependence of hierarchical symmetries on single-electron basis set} \label{sec:sym_analysis}

\begin{figure}[t]
\centering
\includegraphics[width=\columnwidth]{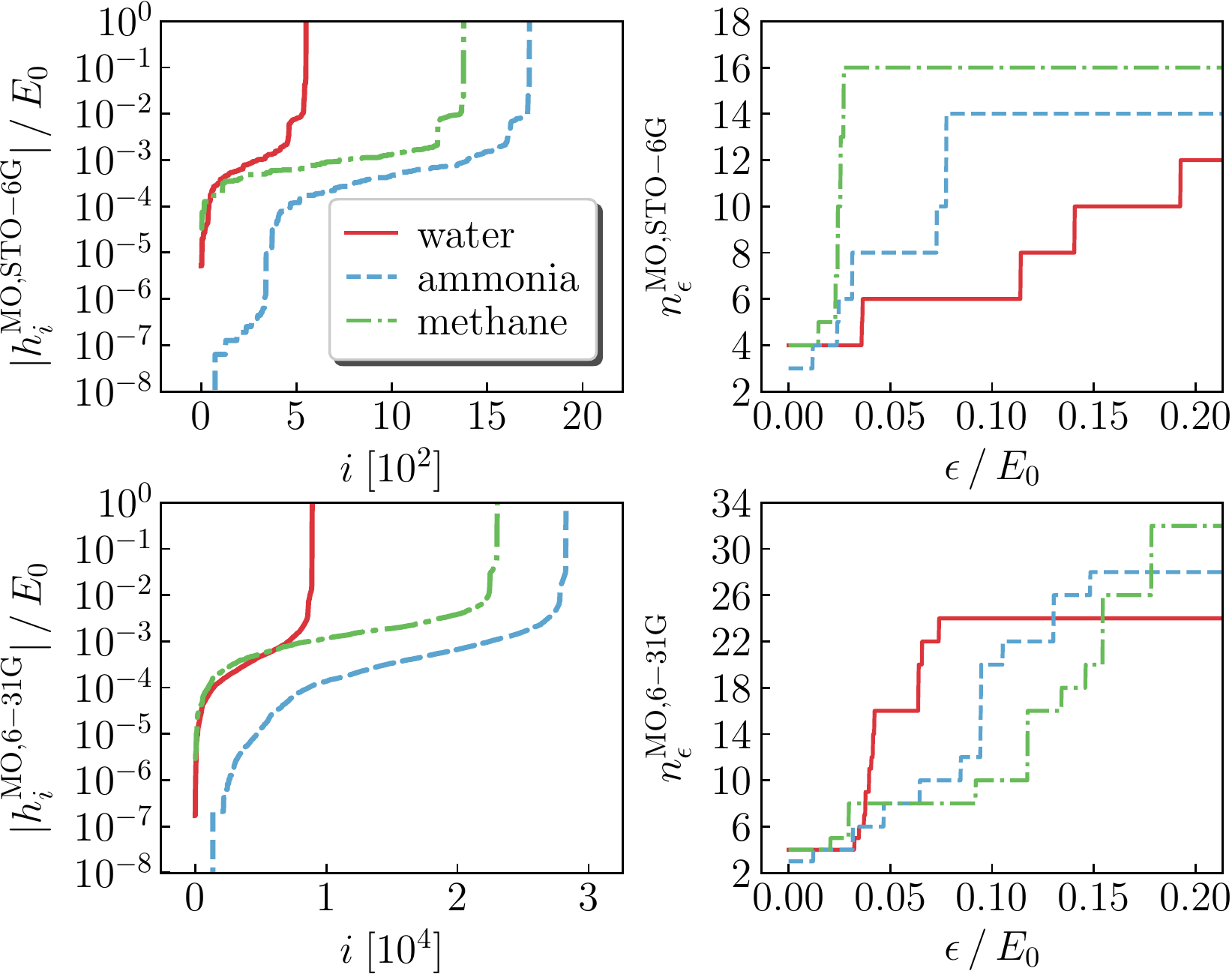}
\caption{Hamiltonian coefficients (left) and number of $\mathbb{Z}_2$ symmetries (right) for water, ammonia, and methane (red circles, blue crosses, green triangles respectively) at STO-6G (top) and 6-31G (bottom) level of theory.}
\label{fig:small_molecules_pople}
\end{figure}

\begin{figure}[t]
\centering
\includegraphics[width=\columnwidth]{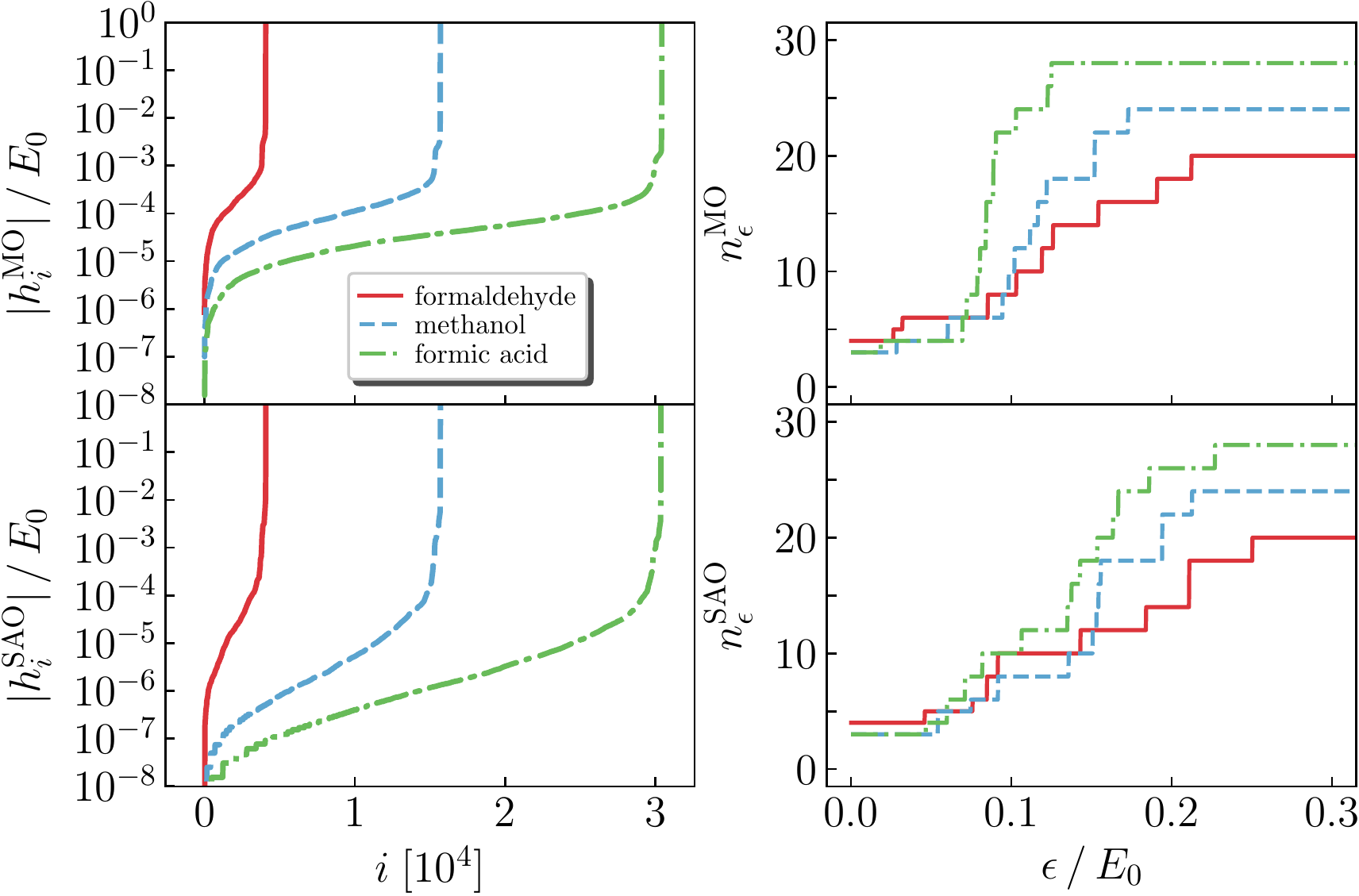}
\caption{Hamiltonian coefficients (left) and number of $\mathbb{Z}_2$ symmetries (right) for formaldehyde, methanol, and formic acid (red circles, blue crosses, green triangles respectively) at STO-6G level of theory, using molecular (top) and symmetrized atomic orbitals (bottom).}
\label{fig:small_molecules_basis}
\end{figure}

The number and structure of the $\mathbb{Z}_2$ symmetries highlighted by the hierarchical procedure may depend on various factors, particularly the single-electron basis on which the second-quantized Hamiltonian is based. The indices $prqs$ in Eq.~\eqref{eq:Helec} label a set of orthonormal orbitals, which are linear combinations of atomic orbitals from an underlying basis set, $\varphi_p({\bf{x}}) = \sum_{\mu} c_{\mu p} \chi_\mu ({\bf{x}})$, where we define a point in space as $\bf{x} \in \mathbb{R}^3$. Both the basis set $\{ \chi_\mu \}$ and the expansion coefficients $c_{\mu p}$ are variables, that can affect
the number of $\mathbb{Z}_2$ symmetries.  It is therefore important to assess the robustness of the  HCT construction as both the basis set and the coefficients $c$ are changed.

Figure~\ref{fig:small_molecules_pople} considers the water, ammonia, and methane molecules at their experimental equilibrium geometries~\cite{johnson2006nist}. We use the minimal STO-6G~\cite{hehre1969self} and the 6-31G basis sets~\cite{ditchfield1971self}, and encode molecular spin-orbitals (i.e. eigenfunctions of the Fock operator) onto qubits using the Jordan-Wigner mapping. In the left portion of the figure, we show the rescaled Hamiltonian coefficients $h_i/E_0$, with $E_0 = \max_i |h_i|$, sorted in ascending order of magnitude.
In the right portion of the figure, we show the number $n_\epsilon$ of $\mathbb{Z}_2$ symmetries of $H_{\epsilon}$. As seen, in both bases the Hamiltonian coefficients retain the same structure, characterized by a smooth decrease in the values of the coefficients at $\epsilon \simeq 10^{-2} E_0$.

\begin{figure*}[t]
  \begin{center}
  \includegraphics[width=\textwidth]{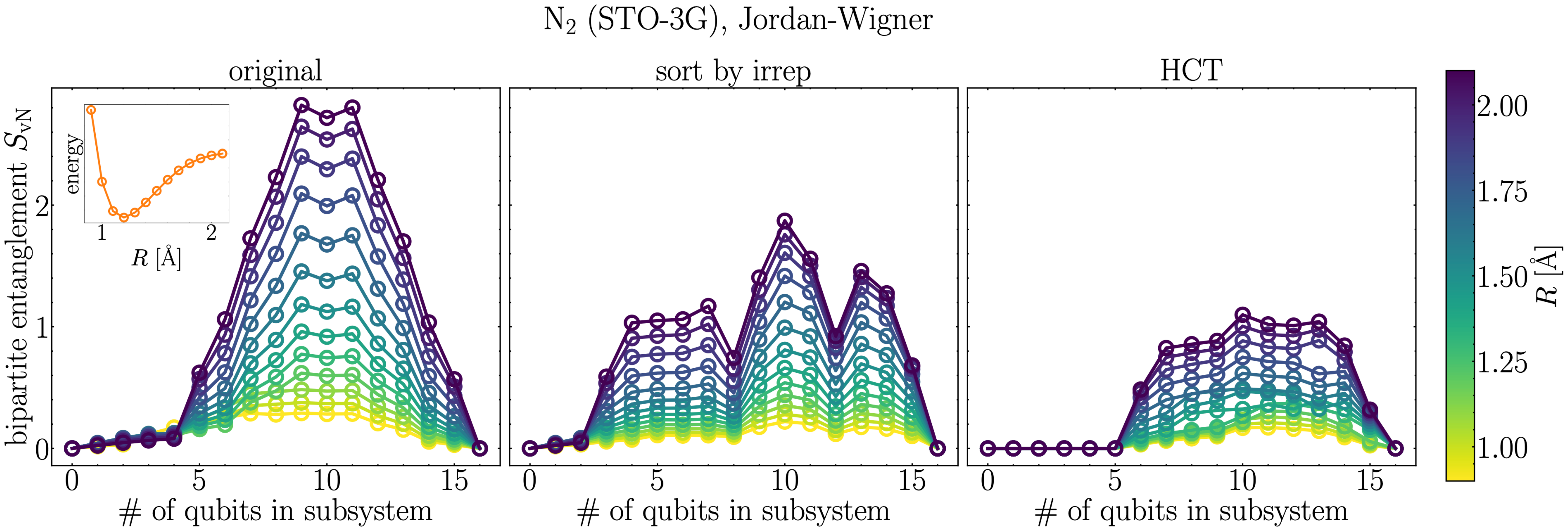}
  \end{center}
  \caption{
  Bipartite von Neumann entanglement entropy versus subsystem size for the ground state of $\mathrm{N}_2$ represented in the STO-3G single-electron basis under Jordan-Wigner qubit mapping for (left) the original energy-based ordering of Hartree-Fock molecular orbitals, (middle) a sorting of molecular orbitals based on their Abelian point group irrep, and (right) the Hamiltonian under HCT, $C^\dagger H C$, as described in the text. The different curves correspond to different bond lengths $R$ ranging from $0.9$\,\AA~to $2.1$\,\AA~in increments of $0.1$\,\AA~as indicated in the color bar; the corresponding ground state energy versus $R$ (dissociation curve) is shown in the inset of the left panel (ranging from -107.677085~Hartree at $R=1.2$\,\AA~to -107.292712~Hartree at $R=0.9$\,\AA). The sharp increase in entanglement as $R$ is increased in the original representation is mitigated most effectively by the HCT scheme, which performs better than the standard irrep-based ordering strategy.
  \label{fig:EE_N2_Rscan}
  }
\end{figure*}

Correspondingly, the number of symmetries $n_\epsilon$ increases from a minimum value $n_\mathrm{sym}$ to a maximum value $n_\mathrm{max}$ when $\epsilon$ is increased from 0 to $\epsilon \simeq 0.2 E_0$. For water, methane, and ammonia (with symmetry groups $C_{2v} \simeq \mathbb{Z}_2 \times \mathbb{Z}_2$, $T_d$ with a subgroup isomorphic to $\mathbb{Z}_2 \times \mathbb{Z}_2$ and $C_{3v}$ with a subgroup isomorphic to $\mathbb{Z}_2$) one has $n_\mathrm{sym}=4,4,2$, respectively. Note that there are some point group symmetries that are not captured by a straightforward application of our method in the standard molecular basis~\cite{setia2020reducing}.  The maximum number of symmetries is $n_\mathrm{max} = 2n_\mathrm{MO}$ (= \# of qubits in standard encodings $=n$) where $n_\mathrm{MO}$ is the number of molecular orbitals. The core molecular orbitals were not included in the simulations. For water, ammonia, and methane this results in $n_\mathrm{MO}=6,7,8$ in a minimal basis and $n_\mathrm{MO}=12,14,16$ in the 6-31G basis.

Figure \ref{fig:small_molecules_basis} considers instead the formaldehyde, methanol, and formic acid molecules at their experimental equilibrium geometries \cite{johnson2006nist}. Using a minimal STO-6G basis for illustrative purposes, we consider encoding molecular and symmetrized atomic orbitals (SAOs). SAOs are constructed by a L\"owdin orthonormalization procedure~\cite{lowdin1970nonorthogonality}. L\"owdin orbitals are projected onto irreps of the molecular point group, and orbitals within each irrep are subsequently localized with a Foster-Boys procedure \cite{boys1960construction}. As seen, the trends observed in the molecular orbital basis are preserved in the SAO basis, confirming the robustness of the appearance of nontrivial approximate binary symmetries in the hierarchical procedure in presence of a molecular point group and of symmetry-adapted orbitals.

\section{Entanglement structure: Implications for classical  and quantum computations} \label{sec:entanglement}

We now investigate the entanglement structure of quantum chemistry ground states represented in the various bases described above. Specifically, we perform exact diagonalization on ($i$) the original qubit Hamiltonian $H$ from Eq.~\eqref{eq:Hpauli} as well as on ($ii$) $H$ transformed by the HCT described in Sec.~\ref{sec:hct} where the threshold schedule $\{\epsilon_m\}$ is taken to be a finely spaced grid up to an $\epsilon_0$ with $n_{\epsilon_0} = n$ (see discussion at the end of Sec.~\ref{sec:hct}); below, we denote the unitary generating this HCT as $C \equiv C(\{\epsilon_m\})$. Given a ground state wavefunction $\ket{\psi}$ in a particular basis, we compute the bipartite von Neumann entanglement entropy
\begin{equation}
    S_{\mathrm{vN},A} = -\mathrm{Tr}\left[\rho_A \ln \rho_A\right]
    \label{eq:SvN}
\end{equation}
for a series of contiguous bipartitions $A|\bar A$, where $A$ contains $1,\dots,n-1$ qubits (size 0 or $n$ subsystems correspond to $S_\mathrm{vN}=0$) and $\rho_A = \mathrm{Tr}_{\bar A}|\psi\rangle\langle\psi|$ is the reduced density matrix for subsystem $A$ (note that $A \cup \bar A$ is the entire system of $n$ qubits). Here, by \emph{contiguous} we mean according to the ordering of qubits in the Hamiltonian being solved (see below).

Bipartite entanglement measures such as Eq.~\eqref{eq:SvN} are extremely important proxies for understanding the cost of both preparing $\ket{\psi}$ on a quantum computer~\cite{vidal_efficient_2003} as well as representing $\ket{\psi}$ on a classical computer using tensor networks states such as a matrix product state (MPS)~\cite{vidal_efficient_2003,schollwock_density-matrix_2011}. For example, at fixed $n$, preparing $\ket{\psi}$ with a local 1D quantum circuit (asymptotically) requires a depth $O(\max(S_{\mathrm{vN}}))$ [where $\max(S_{\mathrm{vN}}) \equiv \max_A(S_{\mathrm{vN},A})$], while representing $\ket{\psi}$ with an MPS requires a bond dimension $\chi = O(\exp \max(S_{\mathrm{vN}}))$ \cite{vidal_efficient_2003,schollwock_density-matrix_2011}. Therefore, if we can manage to find a representation (basis) for the problem such that the ground state exhibits reduced entanglement in this precise sense, this can in principle have a drastic effect on the simulation cost, both for quantum and classical algorithms. Indeed, we find that our hierarchical Clifford transformation rather dramatically achieves this feat, at least for small test molecules in minimal single-electron bases.

\begin{figure}[t]
  \begin{center}
  \includegraphics[width=0.8\columnwidth]{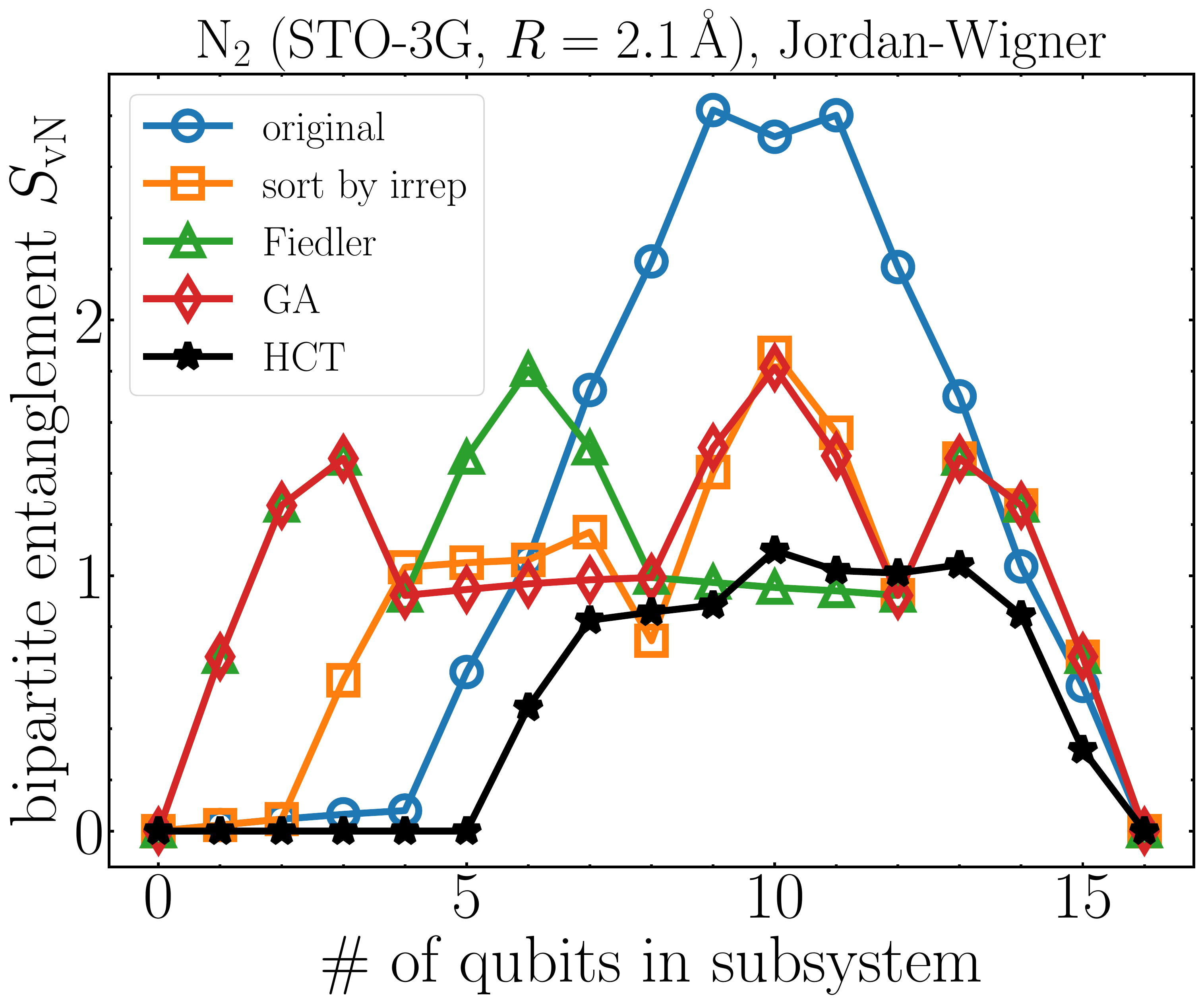}
  \end{center}
  \caption{
  Bipartite von Neumann entanglement entropy versus subsystem size for the same $\mathrm{N}$ shown in Fig.~\ref{fig:EE_N2_Rscan} but only at the largest bond length $2.1$\,\AA (the configuration exhibiting the most entanglement), showing data also for the Fiedler vector and genetic algorithm (GA) orbital ordering strategies. The curves labeled `original', `sort by irrep', and `HCT' are the same $R=2.1$\,\AA data shown in Fig.~\ref{fig:EE_N2_Rscan}. We see that the Fiedler vector and genetic algorithms fail to lower the entanglement below that of the irrep-based ordering strategy.
  \label{fig:EE_N2_fixedR}
  }
\end{figure}

As a first benchmark, we consider stretching diatomic nitrogen $\mathrm{N}_2$ in the minimal STO-3G basis. Freezing the two (out of 10) lowest Hartree-Fock molecular orbitals, results in a problem with $n = 2 \cdot 8 = 16$ qubits. Here we choose to map the electronic Hamiltonian $H_\mathrm{elec}$ [Eq.~\eqref{eq:Helec}] to a qubit Hamiltonian $H$ [Eq.~\eqref{eq:Hpauli}] via the Jordan-Wigner transformation where the spin quantum numbers are taken to be `interleaved' in the mapping to qubits, i.e., even (odd) qubit indices encode occupation of spin $\alpha$ ($\beta$). As is well known in the context of the density matrix renormalization group (DMRG) applied to such quantum chemical systems~\cite{olivares-amaya_ab-initio_2015}, the mere \emph{ordering} of the chosen molecular orbitals can have a drastic effect on $\max(S_{\mathrm{vN}})$ and thus on the DMRG simulation runtime: $O(\chi^3) = O(\mathrm{poly} \exp \max(S_{\mathrm{vN}})) = O(\exp \max(S_{\mathrm{vN}}))$, where $\chi$ is the largest MPS bond dimension used in the DMRG simulation, chosen in order to fix the energy accuracy given the $\max(S_{\mathrm{vN}})$ of the problem. Thus, we consider various state-of-the-art orbital ordering strategies when computing the entanglement entropy for the ground state of the untransformed qubit Hamiltonian $H$. For systems with point group symmetry, it is fruitful to sort the orbitals by their underlying {irreducible representation (irrep) and place bonding orbitals near their antibonding counterparts. For the diatomic molecule we use the D$_{2h}$ Abelian subgroup and sort the irreps according to the order A$_g$, B$_{1u}$, B$_{3u}$, B$_{2g}$, B$_{2u}$, B$_{3g}$, B$_{1g}$, and A$_u$~\cite{ma2013assessment}. Finding an optimal ordering in the general case is NP-hard, and other strategies often involve solving for an approximate solution of a particular NP-hard optimization problem. Here, we focus on the Fiedler vector and genetic algorithm strategies applied to the exchange integral matrix $K_{ij}$ as discussed in Sec.~II E of Ref.~\cite{olivares-amaya_ab-initio_2015} and implemented in the DMRG code \textsc{block2}~\cite{zhai2021low}.

In Fig.~\ref{fig:EE_N2_Rscan}, we compute the bipartite von Neumann entanglement entropy [Eq.~\eqref{eq:SvN}] for all contiguous bipartitions across the $n$ (ordered) qubits of the system while varying the $\mathrm{N}_2$ bond length $R$ from $0.9$\,\AA~to $2.1$\,\AA~in increments of $0.1$\,\AA~(see inset of the left panel for the associated ground state energies). In the three panels, we show $S_{\mathrm{vN},A}$ versus subsystem size for the ground state of the qubit Hamiltonian obtained from the `original' energy-based ordering of the Hartree-Fock orbitals (left, `original'), the qubit Hamiltonian obtained from sorting the orbitals according to their irrep labels (in D$_{2h}$ for $\mathrm{N}_2$) before Jordan-Wigner (middle, `sort by irrep'), and $C^\dagger H C$ (right, `HCT') with $H$ the qubit Hamiltonian in the left panel (i.e.~`original'~\footnote{We find that the ordering of orbitals in $H$ has only a minor (but finite) effect on the entanglement of the ground state of $C^\dagger H C$.}). As $R$ is increased, the entanglement increases in all cases (due to the increased multi-reference character in the underlying ground state wavefunction). While this increase as seen in the left panel is mitigated to a decent extent by using the irrep ordering strategy in the middle panel, the HCT problem representation remarkably gives rise to significantly reduced entanglement still. A few words are in order regarding the qubit ordering strategy used in the HCT plot itself: The leftmost 5 qubits correspond to exact symmetry qubits (see Sec.~\ref{sec:z2review}), while qubits corresponding to approximate symmetries are added sequentially as the threshold $\epsilon$ is increased; i.e., as we proceed to the right in Fig.~\ref{fig:EE_N2_Rscan} (right panel), the $\mathbb{Z}_2$ symmetries represented by their respective `symmetry qubits' on the horizontal axis become less and less exact. While this seems like a natural qubit ordering, there may be even more efficient orderings to reduce the entanglement further still, a topic we leave for future work.

\begin{figure}[t]
  \begin{center}
  \includegraphics[width=0.8\columnwidth]{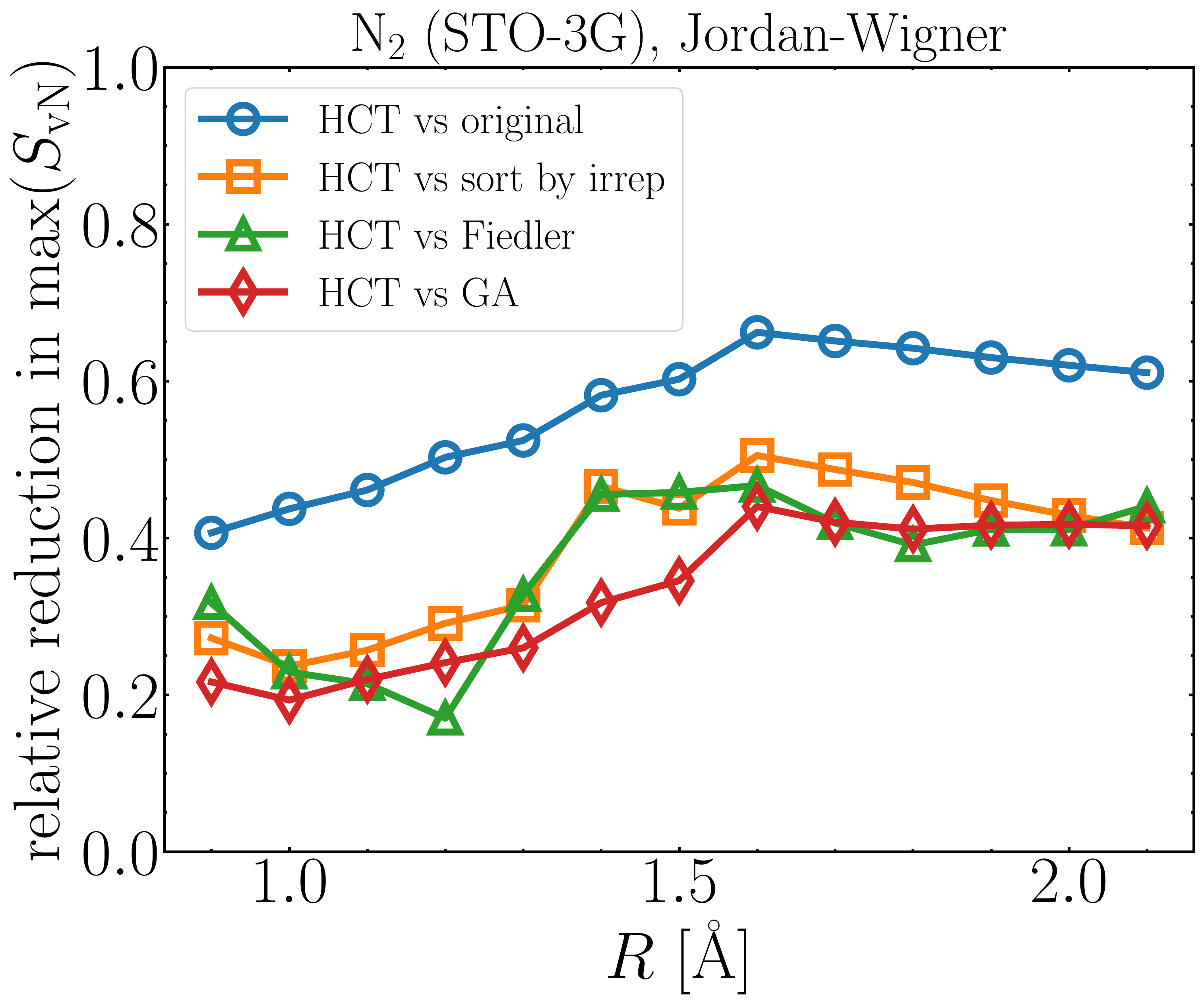}
  \end{center}
  \caption{
  \label{fig:EE_N2_reduction}
  Relative reduction in $\max(S_\mathrm{vN})$ using the HCT representation compared to the other considered strategies (see Figs.~\ref{fig:EE_N2_Rscan} and \ref{fig:EE_N2_fixedR}). (For example, 0.4 implies a $40\%$ reduction.) The gains in entanglement reduction are most robust around $R=1.6$\,\AA, well beyond the equilibrium configuration.
  }
\end{figure}

\begin{figure*}[t]
  \begin{center}
  \includegraphics[width=0.8\textwidth]{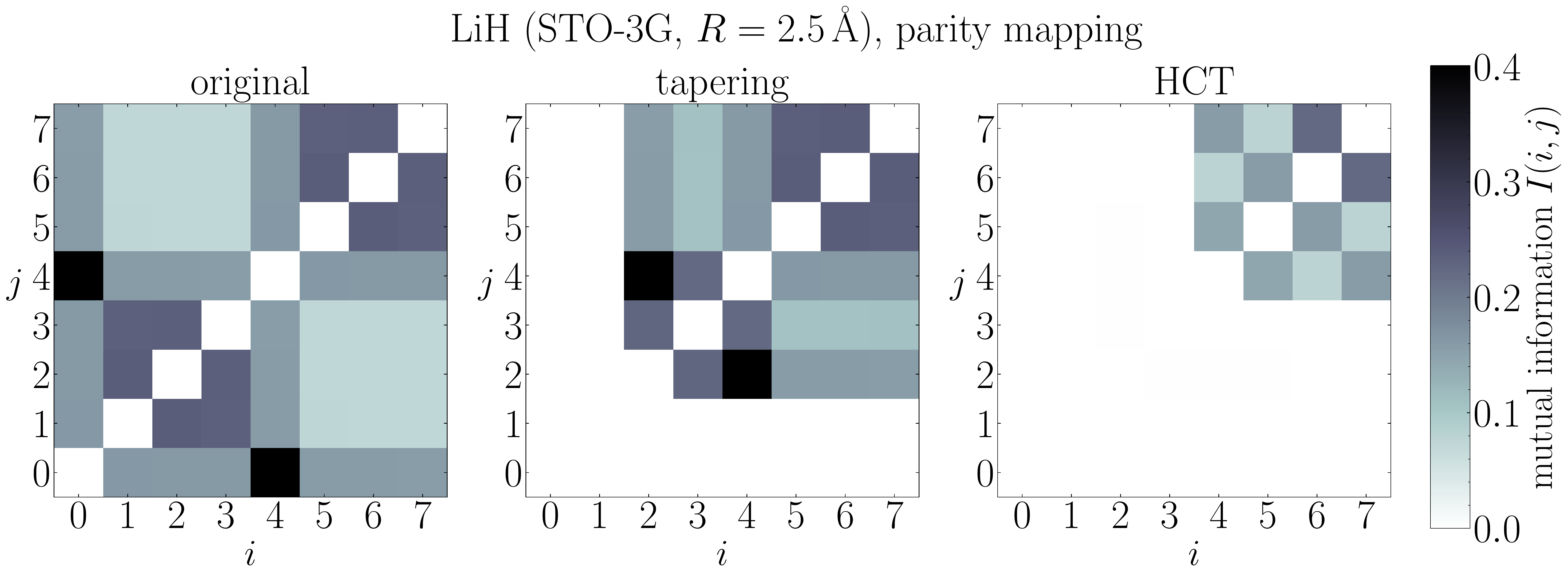}
  \end{center}
  \caption{
  Quantum mutual information between qubits $i$ and $j$, $I(i,j)$, for the ground state of LiH in the STO-3G single-electron basis at (stretched) bond length $R=2.5$\,\AA~under parity qubit mapping for three different Hamiltonian representations: (left) the original qubit Hamiltonian $H$, (middle) the `tapering' representation $C_\mathrm{tapering}^\dagger H C_\mathrm{tapering}$, and (right) the HCT representation $C^\dagger H C$. Two qubits are removed by tapering~\cite{bravyi_tapering_2017}, hence the zeroed out `empty' regions in the middle panel, while two additional nearly exact symmetries are identified by the HCT algorithm. This is the problem setup used for the (hardware-efficient ansatz HCT-based VQE) data shown in the bottom panel of Fig.~\ref{fig:HWE_results}.
  \label{fig:MI_LiH}
  }
\end{figure*}

In Fig.~\ref{fig:EE_N2_fixedR}, we compare the bipartite entanglement entropy curves for also the Fiedler vector and genetic algorithm (GA) strategies at the longest considered bond length $R = 2.1$\,\AA. For $\mathrm{N}_2$, these advanced algorithms do not seem to improve upon the irrep strategy at all, at least with respect to $\max(S_{\mathrm{vN}})$. Finally, in Fig.~\ref{fig:EE_N2_reduction}, we show the relative reduction in $\max(S_{\mathrm{vN}})$ using HCT versus all considered ordering strategies as $R$ is increased~\footnote{For this plot, we show HCT data obtained by conjugating the respective \emph{ordered} $H$ by $C$.}. For example, at $R = 2.1$\,\AA, HCT improves upon the original ordering by more than 60\% and the other strategies by more than 40\%. As we discuss further in the concluding section (Sec.~\ref{sec:conclusion}), if such reductions in $\max(S_{\mathrm{vN}})$ were to persist at nontrivial problem sizes, performing DMRG in the basis generated by our HCT could potentially significantly reduce the runtime cost of the algorithm. Note that both finding and applying the HCT transformation is \emph{efficient} (polynomial in $n$ and $N_\mathrm{Pauli}$). Furthermore, the qubit ordering strategy we employ with the HCT, as described in the previous paragraph, is also efficient to determine and does not involve solution of any NP-hard optimization problems.

Another useful view into the entanglement structure of the ground state is provided by the quantum mutual information:
\begin{equation}
    I(i,j) = S_{\mathrm{vN},i} + S_{\mathrm{vN},j} - S_{\mathrm{vN},ij},
    \label{eq:MI}
\end{equation}
where $S_{\mathrm{vN},i}$ ($S_{\mathrm{vN},ij}$) is the single-qubit (two-qubit) entanglement entropy between qubit $i$ (qubits $i$ and $j$) and the rest of the system. In Fig.~\ref{fig:MI_LiH}, we consider LiH in the STO-3G single-electron basis (with one frozen core orbital) under the parity qubit mapping. Under parity mapping the fixed (known) parities of the spin $\alpha$ and $\beta$ electrons are encoded in the states of two qubits which can thus be subsequently removed from the problem (so-called `two-qubit reduction'), thereby resulting in a $n=8$-qubit problem for LiH in STO-3G. We show in Fig.~\ref{fig:MI_LiH} the quantum mutual information matrix $I(i,j)$ for this model corresponding to the ground state of the original qubit Hamiltonian $H$ [Eq.~\eqref{eq:Hpauli}]~\footnote{Here we use a `block' ordering of the spin quantum numbers such that the first (second) half of the qubit register corresponds to spin $\alpha$ ($\beta$); we also keep the energy-based (`original') ordering of the Hartree-Fock molecular orbitals.} % since reordering of qubits merely results in a reshuffling of the elements in $I(i,j)$.}
(left, `original'), the Hamiltonian in the `tapering'~\cite{bravyi_tapering_2017} basis $C_\mathrm{tapering}^\dagger H C_\mathrm{tapering}$ [Eq.~\eqref{eq:Hp}] (middle, `tapering'), and the Hamiltonian under the HCT $C^\dagger H C$ (right, `HCT'). Notably, while tapering removes two qubits from the problem due to exact binary symmetries in the system, the HCT representation remarkably (nearly) disentangles two additional qubits due to the presence of two approximate---albeit nearly exact---binary symmetries detected by our algorithm. In the next section we perform VQE calculations on this LiH model under HCT and show rather dramatically improved performance relative to VQE runs in the original and tapering representations (see Fig.~\ref{fig:HWE_results}), an improvement which can be directly attributed to the disentangling nature of our HCT.

\section{HCT-based variational quantum eigensolver} \label{sec:vqe}

In this section, we describe a hierarchical variational scheme which exploits the ST introduced in Sec.~\ref{sec:hct}. We begin by spelling out the details of the algorithm in general and then specialize it to two important ansatz classes: hardware-efficient ans\"atze~\cite{kandala_hardware-efficient_2017} and the (Clifford conjugated) UCCSD ansatz~\cite{anand_quantum_2022}. We find that the use of the HCT allows us to reach ground states of molecular systems with lower depths in the case of the hardware efficient ansatz and less optimization calls in the case of the UCCSD ansatz. 

% starting with $m=N$ (the full model) we calculate a set of symmetry groups $\{S_m\}_{m=0,\dots,N}$ with representative generators satisfying $\{\tau^{(m)}_j\}_{j=1,\dots,n_{m}} \subseteq \{\tau^{(m-1)}_j\}_{j=1,\dots,n_{m-1}}$. 

\begin{figure*}
  \begin{center}
  \includegraphics[width=0.95\textwidth]{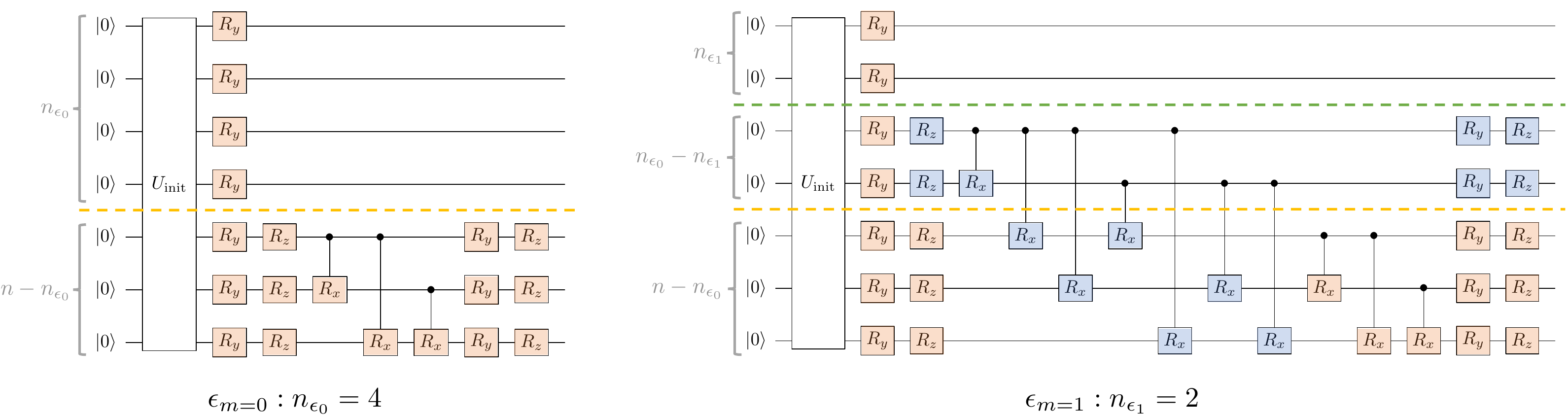}
  \end{center}
  \caption{Illustration of the variational forms for the hardware-efficient HCT-based VQE scheme described in Sec.~\ref{sec:vqe_results}. The total number of qubits is taken to be $n=7$. In the left panel, we show the first step ($m=0$) at threshold $\epsilon_0$ which is assumed to have $n_{\epsilon_0}=4$ symmetries such that $\tilde{H}_{\epsilon_0} = C^\dagger H_{\epsilon_0} C$ is block-diagonal on the $n_{\epsilon_0}=4$ qubits $\{q_{\epsilon_0}(j)\}$; for simplicity, these are taken to be the top $n_{\epsilon_0}=4$ qubits in the register. The symmetry qubits are acted on only by single-qubit $R_y$ gates (to control the symmetry sector), while the remaining `non-symmetry' qubits are entangled with a `depth' $d=1$ `full' (all-to-all) hardware-efficient ansatz. In the right panel, we show the second step ($m=1$) at threshold $\epsilon_1$ now with $n_{\epsilon_1} = 2 < 4$ symmetries, with associated symmetry qubits again taken to live at the very top of the register. Any circuit at $m=0$ can be embedded into this circuit at $m=1$ by zeroing out the blue gates in the latter and matching parameters for the orange gates. In our simulations, we take the initial prepended circuit to be $U_\mathrm{init}|0\rangle^{\otimes n} = C^\dagger|\mathrm{HF}\rangle$.
  \label{fig:hefcirc}
  }
\end{figure*}

Following the recipe spelled out in Sec.~\ref{sec:hct}, given a chosen set of $N$ ordered thresholds $\epsilon_{m=0,\dots,N-1} > 0$ (with $\epsilon_{m+1} < \epsilon_m$ and $\epsilon_N \equiv 0$) and corresponding qubit Hamiltonians [see Eq.~\eqref{eq:Hepsm} and Fig.~\ref{fig:Fig1}], we construct a hierarchical Clifford ST generated by $C \equiv C(\{\epsilon_m\})$~\footnote{Note that in this section, the schedule of $\epsilon_m$ is taken to be some $N = O(5-10)$ points as opposed to the finely spaced grid used in Sec.~\ref{sec:entanglement}.}. We choose to work in the basis implemented by $C$ throughout the rest of the algorithm. To that end, we define a series of transformed truncated Hamiltonians,
\begin{equation}
    \tilde{H}_{\epsilon_{m=0,\dots,N-1}>0} \equiv C^\dagger H_{\epsilon_m} C,
\end{equation}
in addition to the transformed full Hamiltonian,
\begin{equation}
    \tilde{H} = \tilde{H}_{\epsilon_N=0} \equiv C^\dagger H C.
\end{equation}
Focusing on VQE as a solver, we now set out to enhance solution of the latter via iteratively solving the former. While the transformed full Hamiltonian $\tilde{H}$ defines the best variational energy that can be reached by a given ansatz (for a given depth), the use of the series of truncated Hamiltonians can improve the efficiency of optimization to reach this variational minimum.
% The nature of these enhancements depends on the specific form of the ansatz, but conceptually fall into two categories: the use of the Clifford decoupling can lower the required circuit-depth reach a given variational energy, while the definition of a series of Hamiltonians may improve the ability to reach the variational minimum (i.e. improve the optimizability). 

The algorithm begins by solving $\tilde{H}_{\epsilon_0}$ which is block-diagonal on the $n_{\epsilon_0}$ qubits $\{q_{\epsilon_0}(j)\}$. As discussed above, a natural choice for $\epsilon_0$ is some energy scale large enough to give $n_{\epsilon_0} = n$ symmetries (see right panels of Figs.~\ref{fig:small_molecules_pople} and \ref{fig:small_molecules_basis}). It is important to stress that the $n_{\epsilon_0}$ generators themselves used to define $C$ are \emph{not} merely the naive symmetries of $H_{\epsilon_0}$; rather they are `built up' from the full model via increasing $\epsilon$ as described in Sec.~\ref{sec:hct}). 
%Given its relatively large number of symmetries / symmetry qubits, solution of $\tilde{H}_{\epsilon_0}$ is thus relatively `easy'. 
%This relative ease is made clear in Sec.~\ref{sec:vqe_results} where we discuss specific ansatz classes---in general, the required depth/entanglement are substantially reduced when a large number of symmetry qubits are present.
We solve $\tilde{H}_{\epsilon_0}$ using VQE with a circuit ansatz $|\psi_0(\boldsymbol{\theta}_0)\rangle$. Note that we do not do explicit `taper' the model~\cite{bravyi_tapering_2017} at this stage (or any stage throughout the algorithm) as we do not a priori know which symmetry sector to taper into for the ground state of the truncated models; the sector needs to be determined dynamically in the variational optimization procedure. The result of the VQE optimization gives an optimal variational circuit $|\psi_0(\boldsymbol{\theta}_0^*)\rangle$ with energy $E_0^* = \langle\psi_0(\boldsymbol{\theta}_0^*)|\tilde{H}_{\epsilon_0}|\psi_0(\boldsymbol{\theta}_0^*)\rangle$.

Next, we move on to solving $\tilde{H}_{\epsilon_1} = C^\dagger H_{\epsilon_1} C$---which is block-diagonal on $n_{\epsilon_1} \leq n_{\epsilon_0}$ qubits---with a circuit $|\psi_1(\boldsymbol{\theta}_1)\rangle$. We solve this problem via VQE, and very importantly, we initialize the optimization with the circuit obtained for the solution of $\tilde{H}_{\epsilon_0}$: $|\psi_0(\boldsymbol{\theta}_0^*)\rangle$. This amounts to a `warm start', and it requires that all circuits at step $m=0$, $|\psi_0(\boldsymbol{\theta}_0)\rangle$, can be \emph{embedded} into a circuit at step $m=1$ for a specific choice of parameters for the latter. In other words, we must be able to find parameters $\boldsymbol{\theta}_1^\mathrm{init}$ such that
\begin{equation}
    |\psi_1(\boldsymbol{\theta}_1^\mathrm{init})\rangle = |\psi_0(\boldsymbol{\theta}_0^*)\rangle.
    \label{eq:warmstart1}
\end{equation}
If $n_{\epsilon_1} = n_{\epsilon_0}$, this embedding requires no work as the circuit and its block-diagonal structure remains unchanged \footnote{This essentially corresponds to the so-called `hot-starting' procedure of Ref.~\cite{polina_hot-start_2021}.}; however, if $n_{\epsilon_1} < n_{\epsilon_0}$, care must be taken since now the Hamiltonian $\tilde{H}_{\epsilon_1}$ is block-diagonal on less qubits, i.e., the qubits $\{q_{\epsilon_1}(j)\}_{j=1,\dots,n_{\epsilon_1}} \subseteq \{q_{\epsilon_0}(j)\}_{j=1,\dots,n_{\epsilon_0}}$. We describe in detail in Sec.~\ref{sec:vqe_results} how to handle this issue for schemes based on specific ansatz classes.

Minimizing the energy of $\tilde{H}_{\epsilon_1}$ with respect to the circuit $|\psi_1(\boldsymbol{\theta}_1)\rangle$---initialized according to Eq.~\eqref{eq:warmstart1}---produces an optimal circuit $|\psi_1(\boldsymbol{\theta}_1^*)\rangle$ with energy $E_1^* = \langle\psi_1(\boldsymbol{\theta}_1^*)|\tilde{H}_{\epsilon_1}|\psi_1(\boldsymbol{\theta}_1^*)\rangle$. Iterating this procedure another $N-1$ times (for $m=2,\dots,N$), gives a final energy for $\tilde{H}_{\epsilon_N} \equiv \tilde{H} = C^\dagger H C$ of
\begin{equation}
    E_\mathrm{VQE}^* \equiv E_N^* = \langle\psi_N(\boldsymbol{\theta}_N^*)|C^\dagger H C|\psi_N(\boldsymbol{\theta}_N^*)\rangle.
\end{equation}
In the originally defined basis of the full Hamiltonian $H$ [Eq.~\eqref{eq:Hpauli}], the optimal circuit for its ground state is then
\begin{equation}
    |\psi_\mathrm{VQE}\rangle = C|\psi_N(\boldsymbol{\theta}_N^*)\rangle.
\end{equation}
The general embedding condition, cf.~Eq.~\eqref{eq:warmstart1}, reads as follows:
\begin{equation}
    |\psi_m(\boldsymbol{\theta}_m^\mathrm{init})\rangle = |\psi_{m-1}(\boldsymbol{\theta}_{m-1}^*)\rangle,\quad m=1,\dots,N.
    \label{eq:warmstart_gen}
\end{equation}

In the variational context, this procedure seems to takes optimal advantage of the approximate symmetries which are used to build up the HCT described in Sec.~\ref{sec:hct}. In the following  subsection, we describe specific implementations of this HCT-based VQE algorithm based on two specific ansatz classes---hardware-efficient circuits and the (Clifford conjugated) UCCSD ansatz---and present our numerical results based on these two schemes.

\subsection{Variational ansatzae} \label{sec:vqe_results}

The setup for a hardware-efficient implementation (see Fig.~\ref{fig:hefcirc}) is relatively less complicated than for UCCSD (see below), owing---for better or worse---to the the lack of structure characteristic of such variational forms. At the initial step ($m=0$ with highest threshold $\epsilon_{0}$), $\tilde{H}_{\epsilon_0} = C^\dagger H_{\epsilon_0} C$ has $n_{\epsilon_0}$ symmetries which are controlled by $n_{\epsilon_0}$ symmetry qubits $\{q_{\epsilon_0}(j)\}$. Given that the associated symmetry qubit operators are $X$-type Paulis (as we have assumed all along), it suffices to merely apply single-qubit $R_y$ rotations on these $n_{\epsilon_0}$ qubits (with angles $\pm\pi/2$ corresponding to symmetry quantum numbers $\sigma^x_{q_{\epsilon_0}(j)}=\pm1$, assuming initialization in the $\ket{0}$ state) and leave them unentangled from the rest of the system. For the remaining $n-n_{\epsilon_0}$ qubits, we choose some hardware-efficient ansatz: Here we describe a scheme based on the \texttt{EfficientSU2} variational form with controlled-$R_x$ entangling gates which we use in our numerics, although this choice of gate set is obviously not the only possibility.

Each gate is associated with an independent variational angle (not shown in Fig.~\ref{fig:hefcirc}). There is some choice as to which initial state to prepend to the circuit. If solving the full Hamiltonian, it is natural to prepend the Hartree-Fock state and initialize all angles to zero in the variational optimization. However, due to the nature of the truncation in the Pauli basis, it is not straightforward to obtain a Hartree-Fock ground state for the truncated model $H_{\epsilon_0}$. It seems natural however to instead use the Hartree-Fock state for the \emph{full} model, and that is the strategy we take in our numerics below. We must then also apply $C^\dagger$ to the full Hartree-Fock state, which is accounted for by the action of the unitary $U_\mathrm{init}$ in Fig.~\ref{fig:hefcirc}: $U_\mathrm{init}|0\rangle^{\otimes n} = C^\dagger|\mathrm{HF}\rangle$. Initializing in the Clifford-transformed Hartree-Fock state for the full model is not essential however; e.g., for $\epsilon_0$ chosen such that $n_{\epsilon_0}=n$, the ground state of $C^\dagger H_{\epsilon_0} C$ is a single computational basis state (in the local Pauli-$X$ basis) and thus taking $U_\mathrm{init}$ to be the identity suffices equally well. 

At the next step/threshold ($m=1$ with second highest threshold $\epsilon_{1}$) we take the same strategy and apply single-qubit $R_y$ rotations on the $n_{\epsilon_1}$ symmetry qubits $\{q_{\epsilon_1}(j)\}$ and a hardware-efficient circuit on the remaining qubits. This is shown in the right panel of Fig.~\ref{fig:hefcirc}, where we assume $n_{\epsilon_1}=2 (< n_{\epsilon_0} = 4)$.

The embedding condition in Eq.~\eqref{eq:warmstart1} needed for step-to-step initialization is very natural. The optimal circuit obtained in the initial $m=0$ step, i.e, $|\psi_0(\boldsymbol{\theta}_0^*)\rangle$ can be obtained at step $m=1$, i.e., $|\psi_1(\boldsymbol{\theta}_1^\mathrm{init})\rangle$, by (\emph{i}) setting the angles associated with the orange gates in Fig.~\ref{fig:hefcirc} to the their respective optimal values at step $m=0$ and (\emph{ii}) setting the remaining angles (blue gates) to zero ($U_\mathrm{init}$ is held constant throughout). The recipe for an arbitrary step $m$ [see Eq.~\eqref{eq:warmstart_gen}] follows similarly.

Note that the entangling gates must be chosen such that they can be `turned off'; e.g., if instead we used, say, CNOT or `hop' gates in Fig.~\ref{fig:hefcirc}, then we would not in general be able to embed the solution from step $m-1$ into the variational form for step $m$. Also, generalizing (in fact, simplifying) this scheme to `linear' entanglement structure on the non-symmetry qubits is trivial, while `circular'/`sca' entanglement would require a different formulation altogether (if it is possible at all). Extending the scheme to arbitrary depth (repetitions) is also straightforward.

%\subsection{UCCSD scheme} \label{sec:uccsd}

For our UCCSD implementation, we begin by writing the full UCCSD circuit in the form % of a product of $n$-qubit Pauli rotations:
\begin{equation}
    U_\mathrm{UCCSD}(\boldsymbol{\theta}) = \prod_{k=1}^{N_\mathrm{ops}} e^{-i\frac{\theta_k}{2}O_k},
    \label{eq:uccsdfull}
\end{equation}
where $\theta_k \in \mathbb{R}$ and % $P_k \in \mathcal{P}_n$
the $O_k$ consist of a sum of $n$-qubit Pauli operators originating from applying a given qubit mapping to the singles and doubles terms of the UCCSD ansatz~\cite{barkoutsos_quantum_2018}. % These Pauli rotations correspond to Trotterized expansions of the exponentiated singles and doubles terms of the UCCSD ansatz~\cite{barkoutsos_quantum_2018} (note that when written in this form, not all $\theta_k$ are independent parameters).
Again, we want to solve the problem in the approximately block-diagonal (HCT-defined) basis using the iterative procedure described above. In this case---unlike for hardware-efficient circuits---the ansatz contains a lot of physical structure. To retain the physical meaning of the UCCSD ansatz, we first conjugate Eq.~\eqref{eq:uccsdfull} by the Clifford unitary $C$:
\begin{equation}
    \tilde{U}_\mathrm{UCCSD}(\boldsymbol{\theta}) \equiv C^\dagger U_\mathrm{UCCSD}(\boldsymbol{\theta}) C = \prod_{k=1}^{N_\mathrm{ops}} e^{-i\frac{\theta_k}{2}\tilde{O}_k},
    \label{eq:uccsdfulltilde}
\end{equation}
where $\tilde{O}_k \equiv C^\dagger O_k C$ are the operators from Eq.~\eqref{eq:uccsdfull} conjugated by $C$. This Clifford conjugated UCCSD ansatz has the same variational optimum as the original UCCSD ansatz, and we will generally simply refer to it as UCCSD. However, the set of Hamiltonians in the HCT solver changes the optimization landscape of the ansatz which can improve the efficiency.

The bulk of the variational form at step $m$ of the HCT-based VQE algorithm will then consist only of those operators in Eq.~\eqref{eq:uccsdfulltilde} \emph{which conserve the symmetries present at that step}, i.e., commute with all $n_{\epsilon_m}$ symmetry qubit operators $\sigma^x_{q_{\epsilon_m}(j)}$:
\begin{equation}
    \tilde{U}_\mathrm{UCCSD}^{(m)}(\boldsymbol{\theta}^{(m)}) = \prod_{k=1}^{N_\mathrm{ops}^{(m)}} e^{-i\frac{\theta_k^{(m)}}{2}\tilde{O}_k^{(m)}},
    \label{eq:uccsdmtilde}
\end{equation}
where
\begin{equation}
    \left\{\tilde{O}_k^{(m)}\right\} = \left\{\tilde{O}_k~\Big|~[\sigma^x_{q_{\epsilon_m}(j)}, \tilde{O}_k] = 0~\forall~j=1,\dots,n_{\epsilon_m} \right\}.  %_{k=1,\dots,N_\mathrm{ops}^{(m)}}
\end{equation}
Again, there is no `tapering' of qubits in the variational form, i.e., we do not replace the $\sigma^x_{q_{\epsilon_m}(j)}$ appearing in $\tilde{O}_k^{(m)}$ with fixed $\pm1$ eigenvalues ($\mathbb{Z}_2$ parity quantum numbers). Instead, as in the hardware-efficient case, we must determine these quantum numbers variationally. To that end, we merely prepend to Eq.~\eqref{eq:uccsdmtilde} single-qubit $R_y$ rotations acting on the symmetry qubits at this step, $\{q_{\epsilon_m}(j)\}_{j=1,\dots,n_{\epsilon_m}}$, such that $R_{y,q_{\epsilon_m}(j)}\left(\phi^{(m)}_{q_{\epsilon_m}(j)}=\pm\pi/2\right)|0\rangle_{q_{\epsilon_m}(j)} = |\pm\rangle_{q_{\epsilon_m}(j)}$. The full circuit at step $m=0,\dots,N$ then reads as follows:
\begin{equation}
    \tilde{U}_\mathrm{UCCSD}^{(m)}(\boldsymbol{\theta}^{(m)}) \left[ \prod_{j=1}^{n_{\epsilon_m}} R_{y,q_{\epsilon_m}(j)}\left(\phi^{(m)}_{q_{\epsilon_m}(j)}\right)\right] U_\mathrm{init}|0\rangle^{\otimes n},
\end{equation}
where $R_{y,q}$ is a single-qubit $y$-rotation acting on qubit $q$. In addition, we again choose to prepend to this entire variational form the Clifford-transformed Hartree-Fock state for $H$, $U_\mathrm{init}|0\rangle^{\otimes n} = C^\dagger|\mathrm{HF}\rangle$, which is kept fixed at every step in the procedure. This set of variational forms clearly satisfies the necessary embedding property [see Eq.~\eqref{eq:warmstart_gen}] as long as the symmetry qubit $R_y$ gates from the first ($m=0$) step, i.e., $R_{y,q_{\epsilon_0}(j)}\left(\phi^{(0)}_{q_{\epsilon_0}(j)}\right)$, remain active (allowed to vary) for all subsequent steps~\footnote{Note that this is also the case for the $R_y-R_z$-based hardware-efficient scheme of Fig.~\ref{fig:hefcirc}. There are other strategies that one can use instead to slightly decrease the number of variational parameters, but for simplicity we focus on this strategy in our numerics.}.

\begin{figure}[b]
  \begin{center}
  \includegraphics[width=1.0\columnwidth]{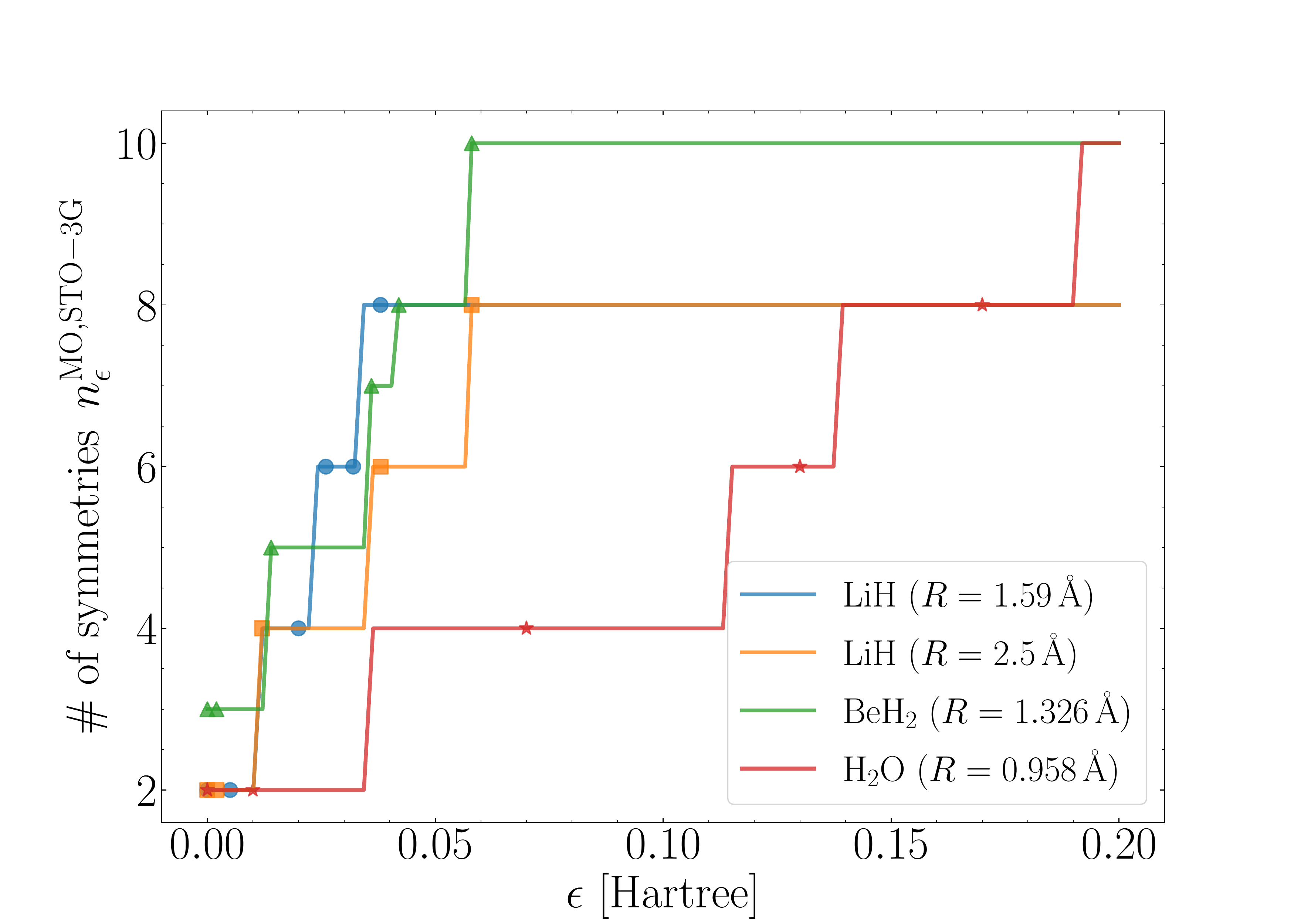}
  \end{center}
  \caption{
  Number of $\mathbb{Z}_2$ symmetries of the truncated Hamiltonian $H_\epsilon$ versus $\epsilon$ for the four molecular systems considered in Figs.~\ref{fig:HWE_results} and Fig.~\ref{fig:UCCSD_results}; this data is analogous to the right columns of Figs.~\ref{fig:small_molecules_pople} and \ref{fig:small_molecules_basis}. The solid points indicate the thresholds $\epsilon_m$ used in our HCT-based VQE scheme (with warm starting).
  \label{fig:vqe_syms_vs_epsilon}
  }
\end{figure}

We now turn to exact statevector simulations (ignoring shot noise and device noise) of the above VQE schemes for several small molecules and assess the accompanying performance improvements relative to more standard methods. Here, we choose to always work with molecules represented in the minimal STO-3G basis, and we employ the parity qubit mapping to arrive at the full multi-qubit Hamiltonian operator $H$; we subsequently eliminate the central and last qubit as afforded by the parity mapping (so-called `two-qubit reduction'---see above). We focus on four specific molecular configurations: LiH at bond lengths $R=1.59$\,\AA~(equilibrium) and $R=2.5$\,\AA~(stretched), BeH\textsubscript{2} at $R=1.326$\,\AA, and H\textsubscript{2}O at $R=0.958$\,\AA. With the chosen single-electron basis and qubit mapping/reduction, LiH corresponds to an $n=8$-qubit problem, while BeH\textsubscript{2} and H\textsubscript{2}O are both $n=10$-qubit problems. In Fig.~\ref{fig:vqe_syms_vs_epsilon}, analogous to the the right panels of Figs.~\ref{fig:small_molecules_pople} and \ref{fig:small_molecules_basis}, we plot the number of $\mathbb{Z}_2$ symmetries at truncation level $\epsilon$ versus $\epsilon$ for these four particular molecular configurations.

% \begin{figure}[h!]
% \centering
% \includegraphics[width=\columnwidth]{vqe_species.pdf}
% \caption{Hamiltonian coefficients (left) and number of $\mathbb{Z}_2$ symmetries (right) for the species studied
% by VQE calculations in the present work, using molecular orbitals from a minimal STO-6G basis.
% BeH$_2$, H$_2$O LiH at equilibrium and stretched bondlength 
% are marked with red circles, blue crosses, green triangles, and orange plus symbols respectively.}
% \label{fig:vqe_syms_vs_epsilon}
% \end{figure}

We present results of the hardware-efficient scheme run on the 8-qubit LiH system (see Fig.~\ref{fig:HWE_results}), as well as the UCCSD scheme run on the the 10-qubit BeH\textsubscript{2} and H\textsubscript{2}O systems (see Fig.~\ref{fig:UCCSD_results}). The selected sets of truncation thresholds used to generate the Clifford STs for each of the four considered molecules/configurations are indicated as solid points on the respective curves of Fig.~\ref{fig:vqe_syms_vs_epsilon}; these are the same thresholds used to run the iterative HCT-based algorithm. In Figs.~\ref{fig:HWE_results} and \ref{fig:UCCSD_results}, we show the energy relative to the exact CI energy, $E - E_\mathrm{exact}$, versus the count of energy evaluations throughout training \footnote{That returned by Qiskit's callback functionality.} for several different strategies.

Results of the last iteration, i.e., that solving the full untruncated model, of the iterative HCT-based VQE algorithm is labeled `HCT (warm start)' and shown in blue; the energy at the beginning of the optimization of this final iteration is plotted as a black diamond in the figures, which in some cases is already at or near `chemical precision' due to the previous iterations. One does not necessarily need to run the iterative VQE algorithm with warm starting in order to take advantage of the employed hierarchical Clifford ST and its accompanying approximately block-diagonal basis. Instead, we can consider running VQE on the full transformed Hamiltonian $C^\dagger H C$ using the variational form from the last step of the iterative procedure, but without the preceding series of warm starts; results of this strategy are labeled `HCT (no warm start)' and shown in orange in Figs.~\ref{fig:HWE_results} and \ref{fig:UCCSD_results}. Additionally, we compare to solving the problem in the `tapering' basis of Ref.~\cite{bravyi_tapering_2017}, i.e., solving $C_\mathrm{tapering}^\dagger H C_\mathrm{tapering}$ of Eq.~\eqref{eq:Hp}, but without a priori tapering the $n_\mathrm{sym}$ symmetry qubits corresponding to exact symmetries into a specific sector (so as to keep the number of qubits the same as in the other strategies); this data is labeled `tapering' and shown in green. Finally, we run traditional VQE on the full model in the original qubit basis (`original'; red).

\begin{figure}[t]
  \begin{center}
  \includegraphics[width=1.0\columnwidth]{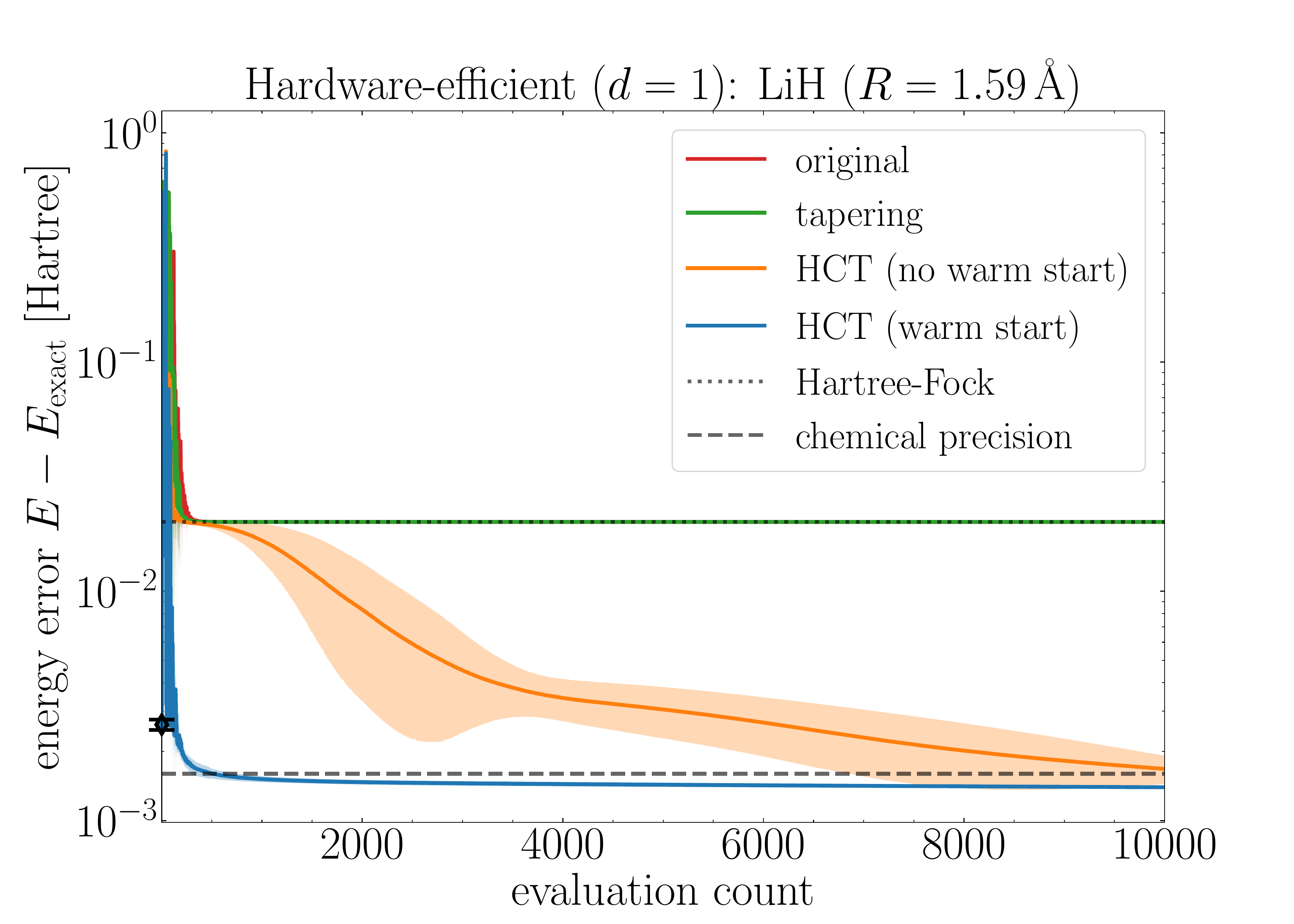}
  \includegraphics[width=1.0\columnwidth]{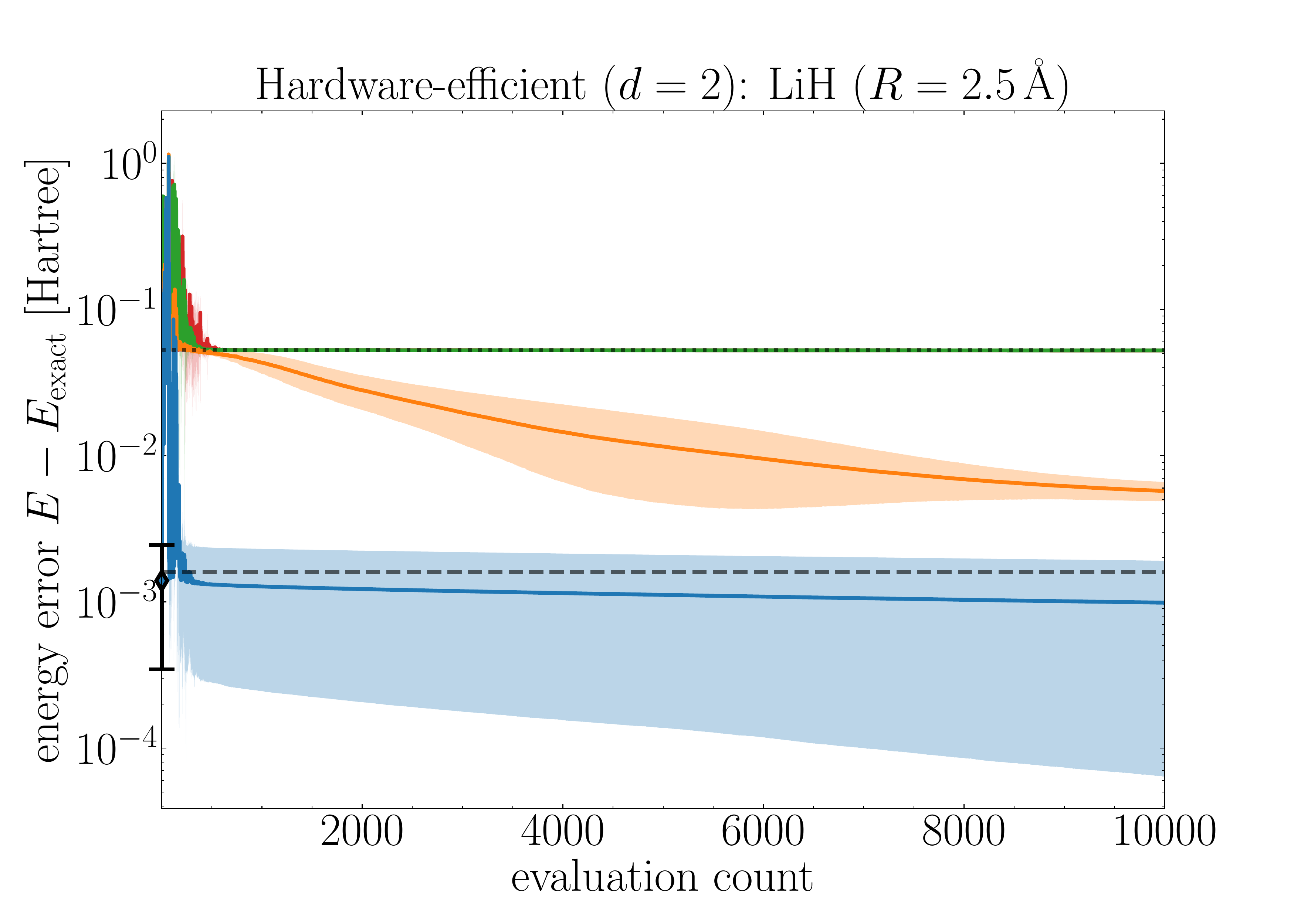}
  \end{center}
  \caption{
  Variational energy versus energy evaluation count during optimization (training trajectories) for hardware-efficient-based VQE simulations (statevector without noise) of LiH in STO-3G ($n=8$ qubits; see text for details). Top: $d=1$ repetition circuits at equilibrium bond length $R=1.59$\,\AA; bottom: $d=2$ repetition circuits at stretched bond length $R=2.5$\,\AA. In each panel, we show VQE runs using the original qubit Hamiltonian representation $H$ (`original'), the Hamiltonian in the $\mathbb{Z}_2$ tapering basis~\cite{bravyi_tapering_2017} $C_\mathrm{tapering}^\dagger H C_\mathrm{tapering}$ (`tapering), and our HCT-based VQE with and without warm starting the solution from threshold to threshold (see text). %[i.e., in the former case we run ordinary VQE on $C^\dagger H C$ with $C \equiv C(\{\epsilon_m\})$].
  For the HCT-based calculations with warm starting, we only show the trajectory for the final step (solution of $H_{\epsilon_N=0} = H$); the energy at the start of the final step optimization is plotted as a black diamond. Here we use the COBYLA optimizer and the shaded regions indicate spread of training trajectories starting from randomly chosen initial parameter points near the Hartree-Fock state (see text for details). Only the HCT-based strategies are able to improve upon the Hartree-Fock energy and are able to achieve `chemical precision' in both cases.
  \label{fig:HWE_results}
  }
\end{figure}

\begin{figure}[t]
  \begin{center}
  \includegraphics[width=1.0\columnwidth]{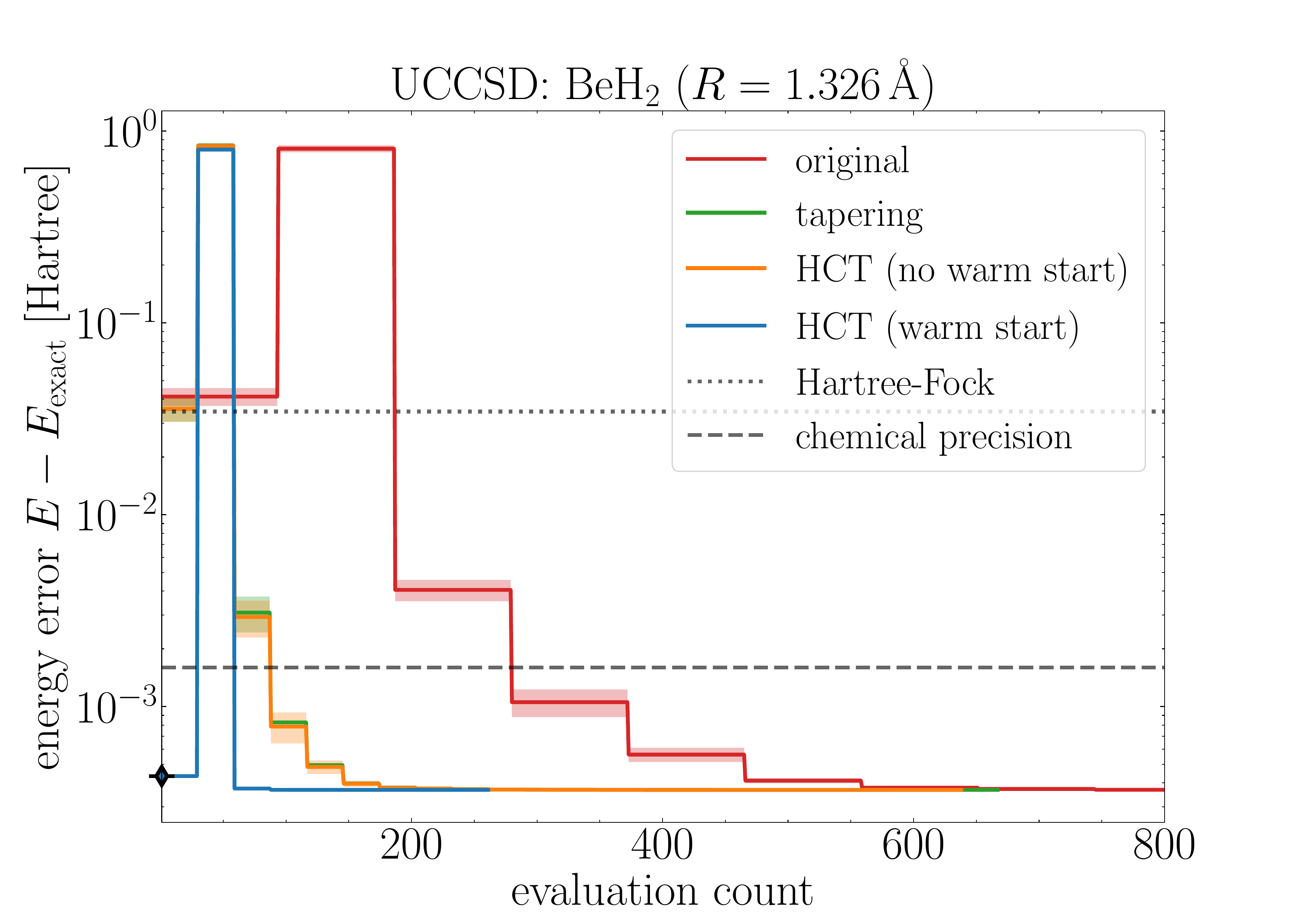}
  \includegraphics[width=1.0\columnwidth]{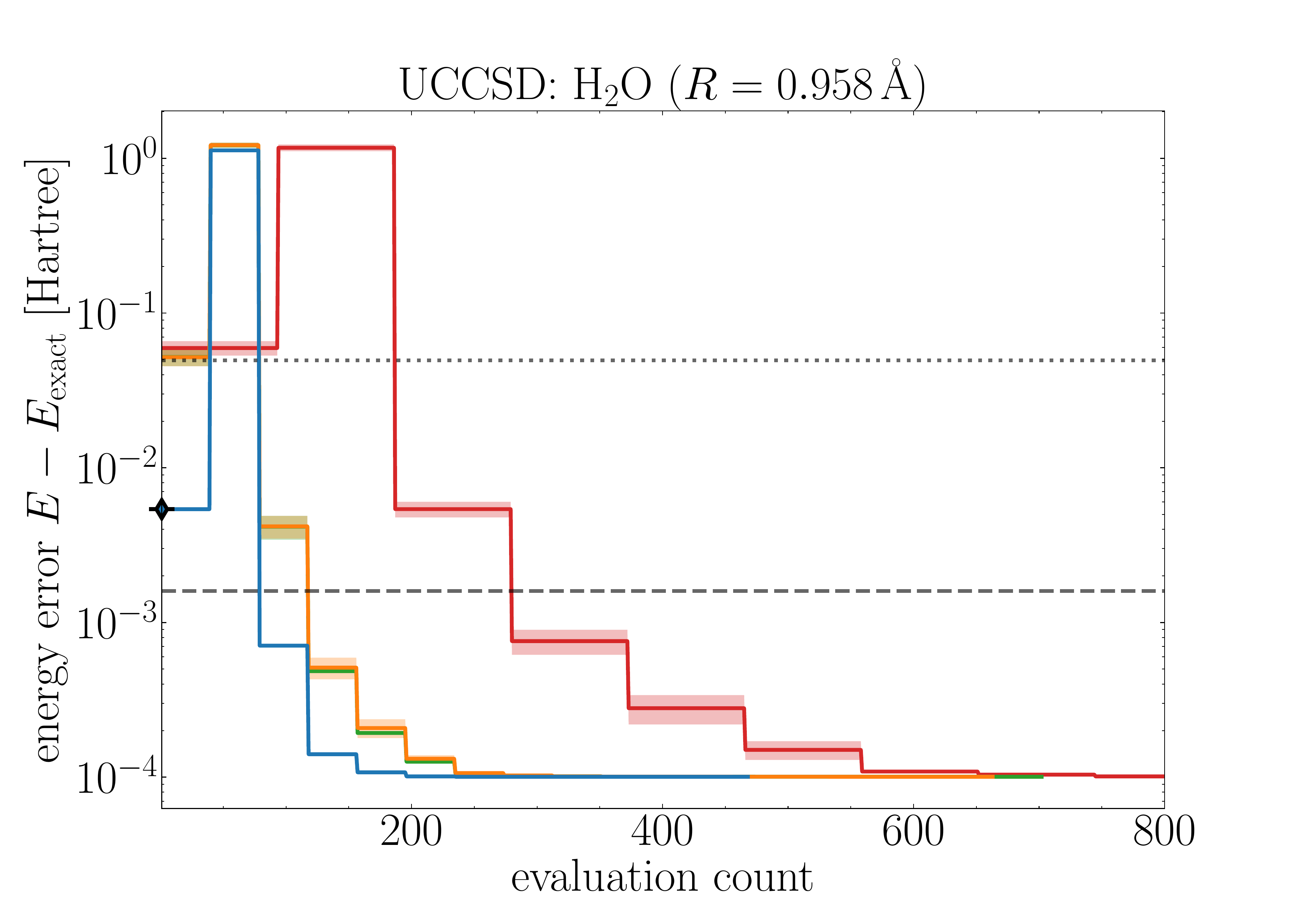}
  \end{center}
  \caption{
  Data analogous to Fig.~\ref{fig:HWE_results} for UCCSD-based VQE simulations of BeH\textsubscript{2} at $R=1.326$\,\AA~(top) and H\textsubscript{2}O at $R=0.958$\,\AA~(bottom), both using the STO-3G single-electron basis ($n=10$-qubit problems; see text for details). Here we choose the L-BFGS-B optimizer. For the HCT-based strategy with warm starting, the training for the final step exhibits relatively fast convergence owing to a good initial trial state obtained by the iterative procedure. %The data sets for all steps of the HCT-based VQE with warm starting shown here and in Fig.~\ref{fig:HWE_results} are presented in Appendix~\ref{app:raw_data}.
  \label{fig:UCCSD_results}
  }
\end{figure}

For all strategies, we consider 10 VQE runs initialized in the Hartree-Fock ground state perturbed slightly by selecting random angles in the variational form drawn from a normal distribution of variance 0.01 (for the iterative strategy, this is done only for the initial $m=0$ step). This has the effect of sending the variational optimization down a spread of training trajectories, the standard deviation of which is indicated by the shaded regions in Figs.~\ref{fig:HWE_results} and \ref{fig:UCCSD_results}. We also indicate the Hartree-Fock energy and `chemical precision' (1.6\,mHartree) with horizontal dotted and dashed lines, respectively, in the figures.
% The raw data for all steps of the iterative algorithm is presented in Appendix \ref{app:raw_data}.

The top panel of Fig.~\ref{fig:HWE_results} shows results for strategies based off a $d=1$ repetition (`depth') hardware-efficient circuit (see Fig.~\ref{fig:hefcirc} for details) applied to LiH at $R=1.59$\,\AA, while the bottom panel shows analogous data for $d=2$ at $R=2.5$\,\AA; here we used the COBYLA algorithm for optimization~\cite{cobyla}. Notably, both traditional strategies---a naive application of VQE in the original qubit basis (`original') and first transforming to the usual `tapering' basis based only on exact $\mathbb{Z}_2$ symmetries~\cite{bravyi_tapering_2017} (`tapering')---fail completely for these Hamiltonians for hardware-efficient circuits at these low depths: the training never improves over the Hartree-Fock energy in either case. On the other hand, the strategies based on the new hierarchical Clifford STs as introduced above (either with or without iterative warm starting) both easily improve over the Hartree-Fock energy, with the iterative strategy achieving `chemical precision' in both cases (for the $d=2$ case at $R=2.5$\,\AA~such iterations are in fact \emph{necessary}, at least within the maximum allowed 10,000 optimizer iterations). This is indeed a dramatic improvement in the performance of relatively low-depth hardware-efficient-circuit-based VQE applied to molecular systems. %In Appendix~\ref{app:raw_data}, we present the raw data for all steps of the iterative warm starting algorithm in Figs.~\ref{fig:LiH_1.59_data} ($d=1$, $R=1.59$\,\AA) and \ref{fig:LiH_2.5_data} ($d=2$, $R=2.5$\,\AA).

Next we consider the UCCSD-based scheme applied to BeH\textsubscript{2} at (symmetric) bond lengths $R=1.326$\,\AA~and H\textsubscript{2}O at $R=0.958$\,\AA. These results are shown in the top and bottom panels of Fig.~\ref{fig:UCCSD_results}, respectively; here, we use the L-BFGS-B optimizer~\cite{lbfgsb}. In this scheme, because we transform both the Hamiltonian and circuit ansatz [see Eq.~\eqref{eq:uccsdfulltilde}], as far as the optimization problem is concerned there is no difference between using the ST based on exact symmetries (`tapering' data) and the new ST based on approximate symmetries in the absence of iterative warm starting [`HCT (no warm start)' data]. Thus the green and orange training curves in Fig.~\ref{fig:UCCSD_results} are equivalent. The influence of the HCT basis change on \emph{depth} of the corresponding (full) UCCSD ansatz when compiled to a native gate set is an interesting detail we leave for future work.
In any case, the full iterative scheme with warm starting does however provide potentially improved convergence properties as the solutions of the truncated Hamiltonians $H_{\epsilon_m}$ with $n_{\epsilon_m} > n_\mathrm{sym}$ involve variational forms with reduced depth and number of parameters. In the examples presented in Fig.~\ref{fig:UCCSD_results}, the data for the last step of the iterative procedure does converge in significantly fewer optimization steps; however, bear in mind that there is a cost associated with getting to the starting points (black diamonds). %as shown in Figs.~\ref{fig:BeH2_data} and \ref{fig:H20_data} of Appendix~\ref{app:raw_data}.
While the whole scheme is very heuristic in nature, it is conceivable that the iterative algorithm could provide substantially improved UCCSD variational performance in terms of, e.g., convergence rate or ease of finding the variational minimum %final energy,
although the selected thresholds, initial state at step $m=0$, and other optimization settings may need to be tuned further.

\section{Conclusion and outlook} \label{sec:conclusion}

The amount of quantum entanglement exhibited by many-body ground states---as characterized quantitatively by measures such as the bipartite entanglement entropy [Eq.~\eqref{eq:SvN}] and quantum mutual information [Eq.~\eqref{eq:MI}]---plays a key role in the difficulty (cost) of finding and representing such ground states using both classical simulation algorithms based on tensor network states, as well as variational quantum algorithms. Clifford transformations, on the other hand, are classically cheap to compute, both when applied as a unitary operation to a computational basis state~\cite{aaronson_improved_2004} as well as when used to conjugate a given operator represented as a sum of Paulis. Indeed Clifford unitaries can themselves exhibit a large degree of entanglement, despite the fact that they can be efficiently computed classically. This suggests a potentially fruitful interplay between appropriately chosen Clifford transformations and `entanglement-limited' classical and quantum many-body algorithms~\footnote{Variational quantum algorithms on near-term devices are mainly limited by entanglement in the sense that large entanglement requires large depth which is prohibitive in the presence of noise.}.

In this work, we have developed a scheme for decoupling quantum degrees of freedom of quantum chemistry ground states through an efficiently computable Clifford transformation which can be applied to a qubit Hamiltonian in preprocessing. To make this notion more precise, if we denote the ground state of the a qubit Hamiltonian $H$ as $|\psi\rangle$ and the ground state of the hierarchical-Clifford-transformed Hamiltonian $C^\dagger H C$ as $|\psi'\rangle$, we find that $|\psi'\rangle$ can have significantly reduced entanglement (see Sec.~\ref{sec:entanglement}). The ground states in the two bases are related as $|\psi\rangle = C|\psi'\rangle$ which---since $|\psi'\rangle$ contains less entanglement than $|\psi\rangle$---can be interpreted to mean that $C$ provides a certain share of the entanglement of $|\psi\rangle$, or equivalently that $C^\dagger$ `disentangles' $|\psi\rangle$ (since $|\psi'\rangle = C^\dagger|\psi\rangle$). 

We stress that our Clifford transformation generated by $C \equiv C(\{\epsilon_m\})$ has clear physical meaning: It identifies approximate binary $\mathbb{Z}_2$ Pauli symmetries in the Hamiltonian via a natural thresholding/decoupling procedure [see Eq.~\eqref{eq:Hepsm}] and transforms the identified $n-n_\mathrm{sym}$ approximate symmetries (in addition to the $n_\mathrm{sym}$ exact binary symmetries~\cite{bravyi_tapering_2017}) to \emph{single-qubit operators}. In other words, the physical meaning of the computational basis states of the qubits takes on an entirely new meaning in the basis defined by our HCT: Instead of encoding information about orbital occupation, as is the case in the original Jordan-Wigner or parity mapped qubit representations, the qubit states now encode the parities of the identified (approximate or exact) binary symmetries~\footnote{Since the `symmetry qubits' representing the identified symmetry generators in the new basis are all $\sigma^x$ operators in the examples considered here, it is the local $\sigma^x$ computational states, $\sigma^x|\pm\rangle = \pm|\pm\rangle$, that encode the parities.}. Therefore, our method should be most effective at easing solution of problems for which many good approximate binary symmetries exist. We note however that our transformation $C$ contains a product of $O(n)$ nontrivial Clifford unitaries and thus has the ability to alter the entanglement structure in the scaling (large $n$) limit~\footnote{This should be contrasted with the original exact symmetry tapering procedure~\cite{bravyi_tapering_2017} which only disentangles $O(1)$ qubits and is thus likely less useful when $n$ is large.}. Thus, we believe the potential is quite high for our Clifford transformation to reduce (classical or quantum) simulation costs of nontrivial quantum chemical systems through entanglement reduction.

In fact, this potential is promising for use in DMRG. Since the runtime cost of DMRG is $O(\exp \max(S_\mathrm{vN}))$, i.e., exponentially sensitive to $\max(S_\mathrm{vN})$, reductions in ground state entanglement such as those shown in Fig.~\ref{fig:EE_N2_reduction} can in principle have a significant effect on the simulation cost of DMRG. For example, a putative $50\%$ reduction in $\max(S_\mathrm{vN})$ as a result of applying the HCT implies a quadratic speedup for DMRG: Using standard arguments~\cite{schollwock_density-matrix_2011}, a $\chi=10,000$ calculation in the original representation can in principle then be done at the same accuracy in the HCT representation using only $\chi=O(100)$ states! We note that within the HCT basis, the Hamiltonian is still a sum of the same number of Pauli strings, and thus the efficiency of a naive term-by-term evaluation of Hamiltonian expectation values within DMRG is unchanged using the HCT basis. However, further investigation is required to understand the complexity of more sophisticated methods to evaluate Hamiltonian expectation values in DMRG, for example through matrix product operators, or how U(1) or SU(2) symmetries can be efficiently implemented in this basis.
%In fact, even modest reductions in entanglement are important: For example, a 10\% reduction would imply a reduction in bond dimension from $\chi=10,000$ to $\chi=(10,000)^{1-0.1} \approx 4000$, still a substantial gain. There are some caveats however. At this point, we are not sure how the HCT affects the MPO bond dimension of the Hamiltonian $C^\dagger H C$ being solved, nor how reasonable it is to utilize U(1) or SU(2) symmetries in a DMRG code solving such a Hamiltonian.

Turning to quantum algorithms, the results in Sec.~\ref{sec:vqe} indicate that VQE can be enhanced by incorporating the HCT procedure into the algorithm, not only via its definition of the problem basis but also through use of a warm-starting procedure that goes hand-in-hand with the construction of the HCT itself. Understanding the extent to which these ideas can contribute to overall improved scalability of VQE at large $n$ remains to be seen, although it seems likely that they merely allow solution of modestly larger systems at a given depth.

Applicability of our Clifford transformation to fault-tolerant quantum algorithms is another interesting area to be explored. Looking ahead, it would be interesting to cost out full Hamiltonian ground state projection via phase estimation and its variants \cite{Wecker2014,Reiher2016PNAS,Elfving2020,Goings2022,lin2020near,PRXQuantum.3.040305} when applied to $C^\dagger H C$. We emphasize again that calculating $C^\dagger H C$ as a sum of Paulis is \emph{classically efficient} and should be achievable at large $n$ and $N_\mathrm{Pauli}$: At each threshold, one has to calculate a basis for the finite-field kernel of a $N_\mathrm{Pauli} \times (2n)$ parity check matrix representing the truncated Hamiltonian and perform an appropriate orthogonalization procedure on the obtained vectors---an overall efficient process well-known in the context of finding stabilizer generators for quantum error correcting codes~\cite{gottesman_stabilizer_1997,bravyi_tapering_2017}.

In all, one ultimate goal of the community is to bring difficult practically relevant quantum chemistry problems to within computational reach using whatever means necessary---classical or quantum hardware or some combination thereof. We hope that ideas such as ours can play a meaningful role in this pursuit going forward.

\acknowledgements

We acknowledge useful exchanges with Bela Bauer, Chun-Fu Chen, and Julia Rice. RVM acknowledges support in part by the National Science Foundation under Grant No.~NSF PHY-1748958. GKC was supported by the US DOE, Office of Science, National Quantum Information Science Research Centers, Quantum Systems Accelerator (QSA). Work by HZ was supported by the US DOE, Office of Science, via Award No. DE-SC0019374.

\bibliography{refs}

\end{document}